\documentstyle[12pt,axodraw,epsf]{article} 
\catcode`@=11
\newcount\@tempcntc
\def\@citex[#1]#2{\if@filesw\immediate\write\@auxout{\string\citation{#2}}\fi
\@tempcnta\z@\@tempcntb\m@ne\def\@citea{}\@cite{\@for\@citeb:=#2\do
{\@ifundefined
{b@\@citeb}{\@citeo\@tempcntb\m@ne\@citea\def\@citea{,}{\bf
?}\@warning
       {Citation `\@citeb' on page \thepage \space undefined}}%
    {\setbox\z@\hbox{\global\@tempcntc0\csname b@\@citeb\endcsname\relax}%
     \ifnum\@tempcntc=\z@ \@citeo\@tempcntb\m@ne
       \@citea\def\@citea{,}\hbox{\csname b@\@citeb\endcsname}%
     \else \advance\@tempcntb\@ne \ifnum\@tempcntb=\@tempcntc
      \else\advance\@tempcntb\m@ne\@citeo
      \@tempcnta\@tempcntc\@tempcntb\@tempcntc\fi\fi}}\@citeo}{#1}}
\def\@citeo{\ifnum\@tempcnta>\@tempcntb\else\@citea\def\@citea{,}%
  \ifnum\@tempcnta=\@tempcntb\the\@tempcnta\else
   {\advance\@tempcnta\@ne\ifnum\@tempcnta=\@tempcntb \else
   \def\@citea{--}\fi
   \advance\@tempcnta\m@ne\the\@tempcnta\@citea\the\@tempcntb}\fi\fi}
   \catcode`@=12

\def\theequation{\arabic{section}.\arabic{equation}}
\begin{document}

\begin{flushright}
MPI/PhT/97--002\\ 
hep-ph/9702393\\ 
January 1997
\end{flushright}

\begin{center}
{\Large {\bf Resonant CP violation induced by particle
  mixing}}\\[0.4cm] 
{\Large {\bf in transition amplitudes }}\\[2.5cm]
{\large Apostolos Pilaftsis}\footnote[1]{E-mail address:
  pilaftsi@mpmmu.mpg.de}\\[0.4cm] 
{\em Max-Planck-Institut f\"ur
       Physik, F\"ohringer Ring 6, 80805 Munich, Germany}
\end{center}
\vskip1.7cm 
\centerline{\bf ABSTRACT} 
We analyze CP violation in resonant transitions involving scalar as
well as fermionic intermediate states, using a gauge-invariant
resummation approach implemented by the pinch technique. We derive the
necessary conditions for resonantly enhanced CP violation induced by
particle mixing, by paying special attention to CPT invariance, and
apply the results of our analysis to two representative new-physics
scenarios: (i) the indirect mixing of a CP-even Higgs scalar with a
CP-odd Higgs particle in two-Higgs doublet models and (ii) the 
CP-asymmetric mixing between the top quark and a new heavy up-type quark,
$t'$, in mirror fermion models.  Furthermore, we explicitly
demonstrate the equivalence of our scattering-amplitude formalism with
that based on the effective Hamiltonian and discuss the possibility of
maximal CP violation in the limiting case of two degenerate
particles.\\[0.3cm] 

\newpage 
\setcounter{equation}{0}
\section{Introduction}

The first experimental observation, which has shown conclusively that
nature laws are not invariant under charge and parity (CP)
transformations, took place in the kaon complex some time ago
\cite{CCFT}. The CP-violating phenomenon observed originates from the
CP-asymmetric particle mixing of a $K^0$ with its CP-conjugate state,
$\bar{K}^0$ \cite{review}. So far, this has been the only example of
CP violation in nature, which has been established experimentally.

One of the most phenomenologically successful theories describing the
time evolution and the mixing of unstable particles is the known
approximation due to Weisskopf and Wigner (WW) \cite{WW}.  In the WW
approximation, an unstable particle system is described through an
effective Hamiltonian and the whole time dependence of the system may
well be determined by an effective Schr\"odinger equation
\cite{LOY}. Another interesting approach is the one formulated by
Sachs \cite{Sachs}, which makes use of the dynamical properties of the
complex pole of the propagator.  However, both formulations mentioned
above are, to some extend, phenomenologically oriented and face
serious difficulties to accommodate important quantum field-theoretic
properties for theories with enhanced predictive power, such as
renormalizable gauge field theories. Apart from developing a
well-defined renormalization scheme to cope with ultra-violet (UV)
divergences at high quantum orders \cite{Ren}, one has to worry that
gauge symmetries \cite{NP,JP&AP} and other required properties, such
as unitarity, analyticity, etc., are preserved within such theories
\cite{JP&AP}.

In our analysis, we shall consider a manifestly gauge-invariant
approach for resonant transitions \cite{JP&AP}, which is implemented
by the pinch technique (PT) \cite{JMC}. This approach is free from
CP-odd gauge artifacts, since it preserves all the discrete symmetries
of the classical action after quantization.  For example, it reassures
the absence of off-mass shell transitions between particles with
different CP quantum numbers in a CP-invariant and anomaly-free
theory.

Over the last years, there has been an increasing interest in CP
asymmetries induced by finite widths of unstable particles
\cite{AP1,AP2,PNsusy,CPetal}.  For example, CP-violating effects may
arise from the interference of the top quark with a new up-type quark
$t'$ in many extensions of the Standard Model (SM) \cite{AP1,AP2} or
from similar interference width effects in the top-quark decay
\cite{CPetal} in a two-Higgs doublet model. In Ref.\ \cite{PNsusy},
the authors considered CP violation through scalar-quark mixing in
supersymmetric (SUSY) models.\footnote[1]{Most recently, we became
  aware of studies, which have extended this idea to the scalar
  lepton sector in SUSY theories \cite{SUSYslept}.} Since SUSY imposes
an approximate mass degeneracy between the scalar strange,
$\tilde{s}$, and down, $\tilde{d}$, quarks \cite{SUSYrel}, it was
found that there is a resonant phenomenon of CP violation due to
$\tilde{s}$-$\tilde{d}$ oscillations, which can be rather large, {\em
  i.e.}, of order 20$\%$. All these CP-violating phenomena may have
dynamical features similar to those of the $K^0\bar{K}^0$ system. We
will exemplify this connection for certain models beyond the SM.

Recently, we have studied resonant CP-violating transitions of a
CP-even Higgs scalar, $H$, into a CP-odd Higgs particle, $A$
\cite{APRL}.  These transitions, which come from a non-vanishing $HA$
mixing, exhibit the very same dynamics known from the $K^0\bar{K}^0$
system \cite{Kabir}.  The size of CP violation has been estimated to
be fairly large, {\em i.e.}, of order one, for a range of kinematic
parameters, preferred by SUSY.

In this paper, we shall present a comprehensive field-theoretic
formalism for resonant CP violation through particle mixing in
scattering amplitudes. The mixed particles, which occur in the
intermediate state as resonances, are either bosons or fermions.  As
will be seen, a resonant enhancement of CP violation can only take
place within this formalism, if at least two of the intermediate
particles are nearly degenerate, {\em i.e.}, their mass difference is
comparable to their widths. This requirement of approximate mass
degeneracy, which appears to be a fine-tuning of the kinematic
parameters, may be the result of some enforced symmetry in the
low-energy limit of the theory, such as SUSY, compositeness, or a
general horizontal symmetry.  Even though we will not specify all the
details of our CP-violating models at very high-energies, {\em e.g.},
at the grand unification scale, it is, however, conceivable to assume
that there exist heavy degrees of freedom, which can break such a
low-energy symmetry and hence induce a small splitting in the masses
of the intermediate particles. Consequently, this effect is quite
analogous to what happens in the $K^0\bar{K}^0$ system, in which the
exact mass degeneracy of the strong states $K^0$ and $\bar{K}^0$
becomes an approximate one when the weak interactions are taken into
account.

The paper has the following structure: In Section \ref{sec:Bos}, we
consider the mixing of bosonic particles, where the PT resummation
approach for particle mixing in scatterings is presented.  For
illustration, we assume a $2\to 2$ scattering process involving
resonant transitions of two Higgs scalars $H$ and $A$, with opposite
CP parities: CP$(H)=+1$ and CP$(A)=-1$. If the Higgs particles $H$ and
$A$ mix, then this mixing phenomenon is described by a non-diagonal
$2\times 2$ propagator matrix in the transition amplitude, which
results from summing up a geometric series of $HH$, $AA$ and $HA$
self-energies.  Furthermore, issues of mixing and mass renormalization
in scalar field theories are discussed in Appendices A and B,
respectively.  In Section \ref{sec:RCPV}, we derive the necessary
conditions for resonantly enhanced CP violation in the $HA$ system and
argue that such a mixing-induced CP violation shares common features
with the CP-violating phenomenon through indirect mixing in the kaon
system.

It is known that invariance of all interactions under the combined
action of CP and time reversal (T) is a fundamental property of the
underlying Hamiltonian. Therefore, in Section \ref{sec:CPT}, we
discuss possible constraints on transition amplitudes, which may
originate from CPT invariance. We find that CPT symmetry generally
leads to a modest reduction of the size of the CP asymmetries.

In Section \ref{sec:WW}, we briefly review the effective Hamiltonian
approach and explicitly demonstrate the equivalence of that approach
with our formalism.  Then, we investigate the possibility of maximal
CP violation for the limiting case of two degenerate particles.  We
observe that, if the effective Hamiltonian expressed in a
$K^0\bar{K}^0$-like basis has the Jordan form, the two
mass-eigenvalues of the effective Hamiltonian are {\em exactly} equal
and, in this extreme case, CP violation through mixing takes its
maximum allowed value.

In Section \ref{sec:HA}, we discuss models that predict a potentially
large $HA$ mixing.  Such a mixing can naturally occur within two-Higgs
doublet models either at the tree level, if one adds softly
CP-violating breaking terms to the Higgs potential, or at one loop,
after integrating out heavy degrees of freedom that break the CP
invariance of the Higgs sector, such as heavy Majorana neutrinos. In
Section \ref{sec:CPHA}, we investigate phenomenological examples of
resonant CP violation through $HA$ mixing.  Numerical estimates reveal
that the CP-violating phenomenon could be of order one and may hence
be observed at future high-energy $pp$, $e^+e^-$ or $\mu^+\mu^-$
machines. Bounds from electric dipole moments (EDM's) of the neutron
and electron are also implemented in this analysis.

In Section \ref{sec:Fer}, we examine fermionic mixing, which is more
involved in comparison with that of bosons owing to the spinorial
structure of fermions, resulting in four degrees of freedom, {\em
  i.e.}, left- and right-handed fields for particles and
anti-particles.  As for the phenomenon of CP violation through mixing,
we find that it is conceptually similar to that of bosons. If the mass
difference of the two mixed fermions is of order of their widths, the
CP-violating phenomenon becomes resonant. As an example, in Section
\ref{sec:CPtt'}, we analyze a simple new-physics CP-violating
scenario, which predicts an asymmetric mixing between the top-quark
and a new up-type quark, $t'$.  Resonant CP-violating $tt'$
transitions may be probed at the CERN $pp$ Large Hadron Collider
(LHC). Section \ref{sec:Concl} contains our conclusions.

\setcounter{equation}{0}
\section{\label{sec:Bos} Bosonic case} 

In this section, we shall analyze CP violation induced by the mixing
of two bosons with opposite CP quantum numbers in scattering
amplitudes.  First, we shall focus our attention on an example with
two-scalar mixing and then extend our discussion to scalar-vector
mixing, such as the mixing of a Higgs particle, $H$, with the $Z$
boson. Finally, we will comment on transitions involving two vector
particles with two
different CP quantum numbers.
\begin{center}
\begin{picture}(360,100)(0,0)
\SetWidth{0.8} \ArrowLine(0,80)(30,50)\Text(0,89)[l]{$a$}
\ArrowLine(0,20)(30,50)\Text(0,11)[l]{$b$}
\ArrowLine(130,50)(160,80)\Text(160,89)[l]{$c$}
\ArrowLine(130,50)(160,20)\Text(160,11)[l]{$d$}
\DashLine(30,50)(80,50){5}\Text(55,55)[b]{$H,A$}
\DashLine(80,50)(130,50){5}\Text(105,55)[b]{$H,A$}
\GCirc(30,50){10}{0.5} \GCirc(130,50){10}{0.5} \GCirc(80,50){10}{0.5}
\Text(80,0)[b]{{\bf (a)}}

\ArrowLine(230,64)(265,64)\Text(220,64)[]{$a$}
\ArrowLine(230,36)(265,36)\Text(220,36)[]{$b$}
\ArrowLine(295,64)(330,64)\Text(340,64)[]{$c$}
\ArrowLine(295,36)(330,36)\Text(340,36)[]{$d$}
\GOval(280,50)(28,20)(0){0.5} \Text(280,0)[b]{{\bf (b)}}

\end{picture}\\[0.7cm]
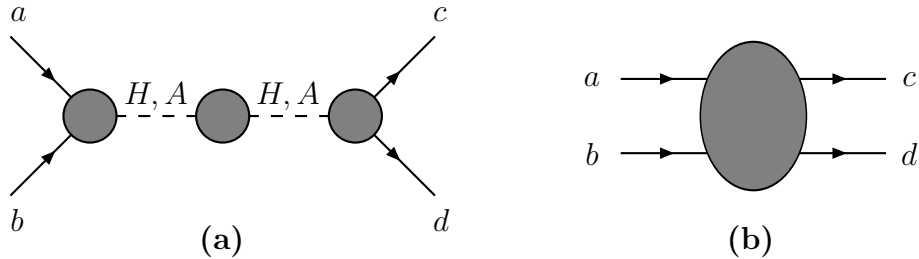
{\small {\bf Fig.\ 1:} Resonant CP violation induced by $HA$ mixing.}
\end{center}

Let us consider the resonant prototype process $ab\to H^*,\, A^* \to
cd$ in Fig.\ 1, where $H$ and $A$ are CP-even and CP-odd (Higgs)
scalars, respectively. The asymptotic states $a,b,c,d$ could be either
fermions, {\em e.g.}, $b$ or $t$ quarks, or vector bosons, {\em e.g},
$W$ or $Z$ bosons.  The transition amplitude of such a process may
conveniently be expressed as
\begin{equation}
\label{Tamp}
{\cal T}\ =\ {\cal T}^{res}\, +\, {\cal T}^{box}\ =\ V^P_i \left(
\frac{1}{s\, -\, {\cal H}(s)}\right)_{ij} V^D_j\ +\ {\cal T}^{box}\, ,
\end{equation}
where 
\begin{equation}
\label{InvDHA}
s\, -\, {\cal H}(s)\ =\ \hat{\Delta}^{-1} (s)\ =\ s\mbox{\bf 1}\, -\,
\left[
\begin{array}{cc}
M^2_A-\widehat{\Pi}^{AA}(s) & -\widehat{\Pi}^{AH}(s)\\
-\widehat{\Pi}^{HA}(s) & M^2_H - \widehat{\Pi}^{HH}(s)
\end{array} \right]\, 
\end{equation}
is the inverse propagator matrix, which describes the dynamics of the
$HA$-mixing system. In fact, the propagator matrix $\hat{\Delta}(s)$
arises from summing up a geometric series of $HH$, $AA$, $HA$ and $AH$
self-energies. In this resummation formalism based on transition
amplitudes, ${\cal H}(s)$ is closely related with the effective
Hamiltonian, obtained in the WW approximation. As will be discussed in
Section \ref{sec:WW}, there is only a minor difference between ${\cal
  H}(s)$ and the effective Hamiltonian. The latter may equivalently be
evaluated from the former, ${\cal H}(s)$, at the resonant region $s\approx
M^2_H\approx M^2_A$, and is therefore $s$-independent.  Furthermore,
$V^P_i$ and $V^D_i$ are the production and decay amplitudes of the
process, as shown in Fig.\ 1(a). In addition, ${\cal T}^{box}$ refers
to the non-resonant part of the amplitude depicted in Fig.\ 1(b), {\em
  i.e.}, $t$-channel or box graphs.

In Eq.\ (\ref{InvDHA}), the symbol hat on the self-energies, {\em
i.e.},  $\widehat{\Pi}^{ij}(s)$ with $i,j=H,A$, has two
meanings. First, it indicates that the diagonal self-energies are
renormalized in some natural scheme, {\em e.g.}, on-mass-shell (OS)
renormalization. In our calculations, we consider the OS scheme, since
it can be implemented much easier than the pole-mass renormalization
scheme.  More details on mixing and mass renormalization in
scalar theories are relegated in Appendices A and B. The off-diagonal
self-energies, $\widehat{\Pi}^{HA}(s)$ and $\widehat{\Pi}^{AH}(s)$,
are UV safe; such CP-violating transitions occur either at the tree
level or are generated radiatively. The second meaning of the symbol
hat refers to the fact that the resummed self-energies should be gauge
independent.  Throughout our analysis, we shall adopt a
gauge-invariant resummation approach implemented by the PT, which
respects the gauge symmetries of classical action at the tree level
\cite{JP&AP}.  The advantages of this method will be seen later on,
when we will discuss the $HZ$ mixing.

The CP-conjugate amplitude, ${\cal T}^{CP}$, may be written down as
follows:
\begin{equation}
\label{TCPamp}
{\cal T}^{CP}\ =\ \overline{{\cal T}}^{res}\, +\, \overline{{\cal
T}}^{box}\ =\ \overline{V}^P_i \left( \frac{1}{s\, -\, \overline{{\cal
H}}(s)}\right)_{ij} \overline{V}^D_j\ +\ \overline{{\cal T}}^{box}\, ,
\end{equation}
where the CP transformation on the production and decay amplitudes may
generally given by
\begin{equation}
\label{Vamp}
V_i^{P,D}\ =\ |V_i^{P,D}|\, e^{i\delta_f}\, e^{i\delta_w}\ \
\stackrel{\displaystyle \mbox{CP}}{\longrightarrow}\ \
\overline{V}_i^{P,D}\ =\ |V_i^{P,D}|\, e^{i\delta_f}\,
e^{-i\delta_w}\, .
\end{equation}
Here, $\delta_f$ denotes the absorptive or final state phase coming
from OS unitarity cuts and $\delta_w$ represents the weak phase.
Under a CP transformation, only $\delta_w$ changes sign. CP-violating
effects in $V_i^{P,D}$ are sometimes called $\varepsilon'$-type
effects in analogy with the $K^0\bar{K}^0$ system. In our study, we
shall ignore that kind of effects, since they are generally small,
unless it is stated otherwise (see also discussion of the fermionic
case in Section \ref{sec:Fer}).  Moreover, we shall neglect the
non-resonant part ${\cal T}^{box}$.  In fact, for sufficiently narrow
resonances, the small value of ${\cal T}^{box} \ll {\cal T}^{res}$ may
be justified near the resonant region. This is equivalent to the
first assumption in the WW approximation \cite{WW}, where direct
transitions between the asymptotic states are omitted. Here, we shall
concentrate on CP-violating effects resulting from the effective
Hamiltonian (mass) matrix, ${\cal H}(s)$.  In a $K^0\bar{K}^0$-like
basis, ${\cal H}(s)$ transforms as
\begin{equation}
\label{CPtrH}
{\cal H}(s)\ \ \stackrel{\displaystyle \mbox{CP}}{\longrightarrow} \ \
\overline{{\cal H}}(s)\, =\, {\cal H}^T(s)\, .
\end{equation}
Note that the effective Hamiltonian in Eq.\ (\ref{InvDHA}) is written
in a different basis. We shall derive the necessary conditions for
resonantly enhanced CP violation in Section \ref{sec:RCPV}.

We must emphasize again that CP violation coming from the
CP-asymmetric mixing in ${\cal H}(s)$ can be much larger than that
from the decay and production amplitudes, $V_i^{P,D}$ and ${\cal
  T}^{box}$.  To quantify the size of CP violation for a generic
transition, ${\cal T}_{FI}$, of some initial state $I$ ({\em e.g.},
$I=a,b$), into the final state $F$ ({\em e.g.}, $F=c,d$), one should
consider the observable
\begin{equation}
\label{aCP}
a_{CP}\ =\ \frac{|{\cal T}_{FI}|^2\, -\, |{\cal
                   T}_{\bar{F}\bar{I}}|^2}{ |{\cal T}_{FI}|^2\, +\,
                   |{\cal T}_{\bar{F}\bar{I}}|^2}\ \approx\
                   \frac{|{\cal T}^{res}|^2\, -\, |\overline{{\cal
                   T}}^{res}|^2 }{ |{\cal T}^{res}|^2\, +\,
                   |\overline{{\cal T}}^{res}|^2}\, ,
\end{equation}
where the state $\bar{I}$ ($\bar{F}$) is the CP transform of $I$
($F$), and hence ${\cal T}^{CP}_{FI}\equiv {\cal
T}_{\bar{F}\bar{I}}$. In Eq.\ (\ref{aCP}), only the resonant
contribution to ${\cal T}$ is taken into account.  For our purposes,
only the $s$-valued matrix $\hat{\Delta}^{-1} (s)$ in Eq.\
(\ref{InvDHA}) needs be inverted. It is then easy to find the entries
of the matrix $\hat{\Delta}(s)$, {\em viz.}
\begin{eqnarray} 
\label{DAA}
\hat{\Delta}_{AA}(s) &=& \Big[ \, s\, -\, M^2_A
+\widehat{\Pi}^{AA}(s)-\, \frac{\widehat{\Pi}^{AH}(s)
\widehat{\Pi}^{HA}(s)}{s-M^2_H+ \widehat{\Pi}^{HH}(s)}\Big]^{-1}\,
,\\
\label{DHH}
\hat{\Delta}_{HH}(s) &=& \Big[ \, s\, -\,
M^2_H+\widehat{\Pi}^{HH}(s)-\, \frac{\widehat{\Pi}^{HA}(s)
\widehat{\Pi}^{AH}(s)}{s-M^2_A+ \widehat{\Pi}^{AA}(s)}\Big]^{-1}\,
,\\
\label{DHA}
\hat{\Delta}_{HA}(s) &=& \hat{\Delta}_{AH}(s)\ =\
-\widehat{\Pi}^{AH}(s) \Big[ \Big(s-M^2_H+\widehat{\Pi}^{HH}(s)\Big)
\Big( s - M^2_A +\widehat{\Pi}^{AA}(s)\Big)\nonumber\\
&&-\widehat{\Pi}^{AH}(s)\widehat{\Pi}^{HA}(s) \, \Big]^{-1}\, .
\end{eqnarray}
Moreover, the Hermiticity condition of the Lagrangian on the real
scalar fields $H$ and $A$ implies that
$\widehat{\Pi}^{AH}(s)=\widehat{\Pi}^{HA}(s)$.

\begin{center}
\begin{picture}(360,200)(0,0)
\SetWidth{0.8} \Vertex(20,150){1.7}
\DashArrowLine(20,150)(54,150){5}\Text(20,159)[l]{$H$}
\Text(20,140)[l]{$p$}
\Photon(64,150)(108,150){3}{5}\Text(115,159)[r]{$Z_{\mu}$}
\GCirc(64,150){10}{0.5}\Text(64,170)[]{$\widehat{\Pi}_{\mu}^{ZH}$}
\Vertex(108,150){1.7} \Text(118,150)[l]{$p^{\mu}$}

\Text(150,150)[l]{$=$}

\Text(200,150)[r]{$i M_Z$} \Vertex(220,150){1.7}
\DashArrowLine(220,150)(254,150){5}\Text(220,159)[l]{$H$}
\DashArrowLine(274,150)(308,150){5}\Text(315,159)[r]{$G^0$}
\GCirc(264,150){10}{0.5}\Text(264,170)[]{$\widehat{\Pi}^{G^0H}$}
\Vertex(308,150){1.7}


\Vertex(20,100){1.7}
\Photon(20,100)(64,100){3}{5}\Text(20,90)[l]{$\mu$}
\Photon(64,100)(108,100){3}{5}\Text(115,90)[r]{$\nu$}
\GCirc(64,100){10}{0.5}\Text(64,120)[]{$\widehat{\Pi}_{\mu\nu}^{ZZ}$}
\Vertex(108,100){1.7} \Text(118,100)[l]{$p^{\mu}p^{\nu}$}

\Text(150,100)[l]{$=$}

\Text(200,100)[r]{$M_Z^2$} \Vertex(220,100){1.7}
\DashArrowLine(220,100)(254,100){5}\Text(220,109)[l]{$G^0$}
\DashArrowLine(274,100)(308,100){5}\Text(315,109)[r]{$G^0$}
\GCirc(264,100){10}{0.5}\Text(264,120)[]{$\widehat{\Pi}^{G^0G^0}$}
\Vertex(308,100){1.7}


\Text(14,30)[r]{$p^{\mu}$} \Vertex(20,30){1.7}
\Photon(20,30)(64,30){3}{5}\Text(40,40)[]{$Z_{\mu}$}
\ArrowLine(64,30)(94,60)\Text(99,60)[l]{$f$}
\ArrowLine(94,0)(64,30)\Text(99,0)[l]{$\bar{f}$}

\Text(150,30)[l]{$=$}

\Text(200,30)[r]{$-i M_Z$} \Vertex(220,30){1.7}
\DashArrowLine(220,30)(264,30){5}\Text(230,39)[l]{$G^0$}
\ArrowLine(264,30)(294,60)\Text(299,60)[l]{$f$}
\ArrowLine(294,0)(264,30)\Text(299,0)[l]{$\bar{f}$}

\end{picture}\\[0.7cm]
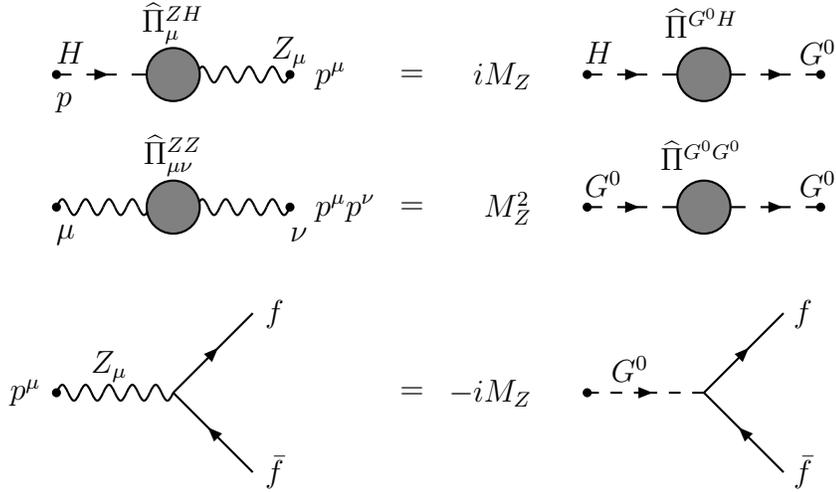
{\small {\bf Fig.\ 2:} PT WI's involving the $HZ$ mixing.}
\end{center}

We will now discuss the mixing of a CP-even (Higgs) scalar, $H$, with
a massive (gauge) vector particle, {\em e.g.}, the $Z$ boson. It is
obvious that the scalar, $H$, can only couple to the longitudinal
component of the gauge particle due to angular momentum
conservation. In spontaneous symmetry breaking (SSB) theories, the
longitudinal component of a gauge boson, $Z$, may be represented
equally well by the respective would-be Goldstone boson, $G^0$, which
is a CP-odd scalar. The advantages of our gauge-invariant resummation
approach may be seen in the description of the $HZ$ system.  In fact,
there are PT Ward identities (WI's) that can be used to convert $ZH$
and $ZZ$ strings into $G^0H$ and $G^0G^0$ ones \cite{PS} before
resummation occurs \cite{JP&AP}. As shown in Fig.\ 2, these identities
are
\begin{eqnarray}
\label{PTWIZH}
p^\mu\widehat{\Pi}_\mu^{HZ}(p)\ -\ i M_Z\widehat{\Pi}^{HG^0}(p^2) &=&
0\, , \qquad p^\mu\widehat{\Pi}_\mu^{ZH}(p)\ +\ i
M_Z\widehat{\Pi}^{G^0H}(p^2)\ =\ 0\, , \nonumber\\ p^\mu
p^\nu\widehat{\Pi}_{\mu\nu}^{ZZ}(p)\ -\
M^2_Z\widehat{\Pi}^{G^0G^0}(p^2) & =& 0\, ,\qquad p^\mu
\Gamma^{Zf\bar{f}}_\mu\ =\ -iM_Z \Gamma^{G^0f\bar{f}}\, .
\end{eqnarray}
Considering the fact that
\begin{equation}
\widehat{\Pi}_\mu^{ZH}(p)\ =\ p_\mu\widehat{\Pi}^{ZH}(p^2)\quad
\mbox{and} \quad \widehat{\Pi}^{ZZ}_{\mu\nu}(p)\ =\
t_{\mu\nu}(p)\widehat{\Pi}^{ZZ}_T(p^2)
+\ell_{\mu\nu}(p)\widehat{\Pi}^{ZZ}_L\, ,
\end{equation}
with $s=p^2$ and
\begin{displaymath}
t_{\mu\nu}(p)\ =\ -g_{\mu\nu} + \frac{ p_\mu p_\nu}{p^2}\, ,\quad
\ell_{\mu\nu}(p)\ = \frac{p_\mu p_\nu}{p^2}\, ,
\end{displaymath}
one obtains the relations
\begin{equation}
p^2 \widehat{\Pi}^{ZH}(p^2)\ =\ -iM_Z\widehat{\Pi}^{G^0H}(p^2)\,
,\quad p^2 \widehat{\Pi}^{ZZ}_L(p^2)\ =\
M^2_Z\widehat{\Pi}^{G^0G^0}(p^2)\, .
\end{equation}
Also, it is important to stress that the vacuum polarizations $\gamma
H$ and $\gamma G^0$ are completely absent within the PT framework
\cite{APRL}, {\em i.e.},
\begin{equation}
\widehat{\Pi}^{\gamma G^0}_\mu(p)\ =\ \widehat{\Pi}^{\gamma H}_\mu
(p)\ =\ 0\, ,
\end{equation}
independently of whether CP violation is present in the theory. This
must be contrasted with the conventional S-matrix prediction in the
$R_\xi$ gauges, in which the $\gamma G^0$ and $\gamma H$ self-energies
do not vanish in general.

Taking the PT WI's in Eq.\ (\ref{PTWIZH}) into account, one then ends
up with a simple coupled Dyson-Schwinger equation system similar to
Eq.\ (\ref{InvDHA}), in which only $H$ and $G^0$ mix. This system may
be described by the following inverse propagator matrix:
\begin{equation}
\label{InvDGH}
\hat{\Delta}^{-1} (s)\ =\ \left[
\begin{array}{cc}
s+\widehat{\Pi}^{G^0G^0}(s) & \widehat{\Pi}^{G^0H}(s)\\
\widehat{\Pi}^{HG^0}(s) & s-M^2_H+\widehat{\Pi}^{HH}(s)
\end{array} \right]\, .
\end{equation}
It is interesting to notice in Eq.\ (\ref{InvDGH}) that the would-be
Goldstone boson, $G^0$, has the desirable property to be massless in
the PT. The inversion of $\hat{\Delta}^{-1} (s)$ proceeds analogously,
by making the identifications $A\equiv G^0$ and $M_A=0$ in Eqs.\
(\ref{DAA})--(\ref{DHA}).

Finally, the mixing of two vector particles, {\em e.g.}, $Z$ and $Z'$,
can be described by an inverse propagator matrix very analogous to
Eq.\ (\ref{InvDHA}), in which only the transverse parts of the vacuum
polarizations $ZZ$, $ZZ'$, and $Z'Z'$ are involved. However, if the
transverse $Z'$ boson is assumed to be CP-odd, it should only have a
coupling of EDM type to fermions. In gauge theories, such a five
dimensional operator cannot be present at the tree level without
spoiling renormalizability.  However, a CP-odd scalar $Z'$ boson can
have tree-level couplings with Higgs scalars \cite{CNP}.  Since the
essential features of a CP-asymmetric mixing between two vector
particles will be identical to those of the scalars, the building of a
particular model that predicts a CP-violating $ZZ'$ mixing may be
studied elsewhere \cite{JR}.

\setcounter{equation}{0}
\section{\label{sec:RCPV} Necessary conditions for resonant CP violation} 

We shall derive the necessary conditions under which CP violation can
be resonantly enhanced in the $HA$ system. We shall perform our
analysis in a $K^0\bar{K}^0$-like basis.  The results of our study may
also carry over to cases involving vector-scalar and vector-vector
particle mixing.

The relation between the $K^0\bar{K}^0$ and $HA$ bases is given by
the following transformations:
\begin{eqnarray}
\label{KKtrafo}
iA &=& \frac{1}{\sqrt{2}}\, \Big(\, K^0\, -\, \bar{K}^0\, \Big)\,
,\nonumber\\ H &=& \frac{1}{\sqrt{2}}\, \Big(\, K^0\, +\, \bar{K}^0\,
\Big)\, ,
\end{eqnarray}
where $\bar{K}^0$ is the Hermitian- and CP-conjugate state of
$K^0$. Expressing the effective Hamiltonian ${\cal H}(s)$ in Eq.\
(\ref{InvDHA}) in the $K^0\bar{K}^0$ basis, we obtain
\begin{eqnarray}
\label{KKHam}
(K^{0*},\bar{K}^{0*})\tilde{{\cal H}}\left(\begin{array}{c} K^0\\
\bar{K}^0
\end{array}\right) & = & \nonumber\\
&&\hspace{-125pt} \frac{1}{2}\, \left[\begin{array}{cc}
M^2_H+M^2_A-\widehat{\Pi}^{HH}-\widehat{\Pi}^{AA} &
M^2_H-M^2_A-\widehat{\Pi}^{HH}+\widehat{\Pi}^{AA}+2i\widehat{\Pi}^{AH}\\
M^2_H-M^2_A-\widehat{\Pi}^{HH}+\widehat{\Pi}^{AA}-2i\widehat{\Pi}^{AH}
& M^2_H+M^2_A-\widehat{\Pi}^{HH}-\widehat{\Pi}^{AA}
\end{array} \right] ,\quad
\end{eqnarray}
where dependence of the self-energies on $s$ is implied.  From Eq.\
(\ref{KKHam}), we see that $\tilde{{\cal H}}$ shares all that
properties known from the kaon system. More explicitly, CPT invariance
requires that
\begin{equation}
\label{CPTHam}
\tilde{{\cal H}}_{11}(s)\ =\ \tilde{{\cal H}}_{22}(s)\, ,
\end{equation}
which holds true in Eq.\ (\ref{KKHam}), as it should. In addition, CP
invariance prescribes the equality
\begin{equation}
\tilde{{\cal H}}_{12}(s)\ =\ \tilde{{\cal H}}_{21}(s)\, .
\end{equation}
This is only valid if $\widehat{\Pi}^{AH}(s)=0$. Thus, the $HA$ mixing
gives rise to CP violation through the effective Hamiltonian (mass)
matrix $\tilde{{\cal H}}(s)$.

As will be seen in Section \ref{sec:WW}, the basic parameter in the WW
approximation that quantifies CP violation through indirect mixing in
the kaon system \cite{review} is given by
\begin{eqnarray}
\label{q/p}
\hspace{-30pt}\Big| \frac{q}{p}\Big|^2 &=& \left|\frac{\tilde{{\cal H}}_{21}}{
\tilde{{\cal H}}_{12}}\right| \ =\ \nonumber\\ &&\hspace{-50pt} \left\{ \frac{
[M^2_H-M^2_A-\Re e(\widehat{\Pi}^{HH}-\widehat{\Pi}^{AA})-2\Im m
\widehat{\Pi}^{HA}]^2+ [\Im m(\widehat{\Pi}^{HH}-\widehat{\Pi}^{AA})
+ 2\Re e\widehat{\Pi}^{HA}]^2}{ [M^2_H-M^2_A-\Re e(\widehat{\Pi}^{HH}
-\widehat{\Pi}^{AA}) +2\Im m\widehat{\Pi}^{HA}]^2+ [\Im
m(\widehat{\Pi}^{HH}-\widehat{\Pi}^{AA})-2\Re e\widehat{\Pi}^{HA}]^2}
\right\}^{1/2}\hspace{-10pt},
\end{eqnarray}
which is a rephasing invariant quantity and hence physically
meaningful.  In the OS renormalization scheme, $\Re e
(\widehat{\Pi}^{HH}-\widehat{\Pi}^{AA})(s)$ is a negligible term, near
the resonant region $s \approx M^2_H \approx M^2_A$, which is formally
of order $g^6$ in the coupling constant. Indeed, expanding the
self-energy $\Re e \widehat{\Pi}^{HH}(s)$ in a Taylor series about
$M^2_H$ yields
\begin{eqnarray}
\label{ReHH}
\Re e \widehat{\Pi}^{HH}(s) & =& \Re e \widehat{\Pi}^{HH}(M^2_H)\,
+\, (s-M^2_H)\Re e \widehat{\Pi}^{HH'}(M^2_H)\nonumber\\
&& +\, \frac{1}{2}\, (s-M^2_H)^2\Re e \widehat{\Pi}^{HH''}(M^2_H)\, 
+\, \dots\, ,
\end{eqnarray}
where the prime denotes differentiation with respect to the variable
$s$. The first two terms on the RHS of Eq.\ (\ref{ReHH}) vanish in the
OS scheme (see also Eqs.\ (\ref{OS1}) and (\ref{OS2}) in Appendix A).
The remaining terms are of order $g^6$ and higher, because $s - M^2_H
= {\cal O}(\Im m\widehat{\Pi}^{HH}(M^2_H)) = {\cal O}(g^2)$. Likewise,
one can show that $\Re e \widehat{\Pi}^{AA}(s)={\cal O}(g^6)$ near the
resonant region of our interest. Therefore, in what follows, we can
safely neglect these high-order terms, after OS renormalization has
been carried out.  It is now instructive to consider the following two
cases:

\begin{itemize}
\item[ (i)] The $HA$ mixing may be predicted at the tree level or
  induced radiatively after integrating out heavy degrees of freedom,
  {\em i.e.}, $\Re e\widehat{\Pi}^{HA}\not= 0$ and it is UV safe.
  Examples of the kind will be discussed in Section \ref{sec:HA}. In
  addition, we assume $\Im m \widehat{\Pi}^{HA} = 0$. Then, in the
  region of interest, $M_H\approx M_A$, the CP-violating mixing
  parameter behaves as
\begin{equation}
\label{q/p1}
\Big| \frac{q}{p}\Big|^2\ \sim \ \left| \frac{ \Im
m(\widehat{\Pi}^{HH}-\widehat{\Pi}^{AA})+2 \Re
e\widehat{\Pi}^{HA}}{\Im m(\widehat{\Pi}^{HH}-\widehat{\Pi}^{AA})-
2\Re e\widehat{\Pi}^{HA}}\right|\ .
\end{equation}
Evidently, if
\begin{equation}
\label{CPN1}
\Im m(\widehat{\Pi}^{HH}-\widehat{\Pi}^{AA})\ \sim \ \pm\, 
                                   2\, \Re e\widehat{\Pi}^{HA}\, ,
\end{equation}
$|q/p|$ takes either very small or very large values, giving rise to
a resonant enhancement of CP violation. Note that for large mass
differences between $H$ and $A$, one has $|q/p|\approx 1$, near the
resonant region, as can be seen from Eq.\ (\ref{q/p}).

\item[(ii)] Another interesting case and, perhaps, equivalent to (i),
  arises when $\Im m \widehat{\Pi}^{HA}\not= 0$ and $\Re e
  \widehat{\Pi}^{HA} = 0$. This possibility may emerge from rotating
  or renormalizing away the finite term $\Re e \widehat{\Pi}^{HA}$ in
  case (i).  For instance, if $\Re e \widehat{\Pi}^{HA}$ is an
  $s$-independent squared mass term and $\Im m\widehat{\Pi}^{HA} =0$,
  one can perform an orthogonal transformation in the Hilbert space
  spanned by the fields $H$ and $A$, such that $\Re e
  \widehat{\Pi}^{HA} = 0$.  This newly defined basis for the fields
  $H$ and $A$ is usually called mass basis. In the mass basis, the
  absorptive self-energy transition $\Im m \widehat{\Pi}^{HA}$ is in
  general different from zero. Note that the transition matrix
  elements ${\cal T}$ and ${\cal T}^{res}$ in Eq.\ (\ref{Tamp}) are
  invariant under such basis transformations of the intermediate
  scalar states and, hence, uniquely determined. For small width
  differences, $\Im m \widehat{\Pi}^{HH}\approx \Im m\widehat{\Pi}^{AA}$, 
  one then finds
\begin{equation}
\label{q/p2}
\Big| \frac{q}{p}\Big|^2\ \sim \ \left| \frac{M^2_H - M^2_A - 2\Im
m\widehat{\Pi}^{HA}}{ M^2_H - M^2_A + 2\Im m\widehat{\Pi}^{HA}}\right|
\ ,
\end{equation}
and the condition for resonant CP violation reads
\begin{equation}
\label{CPN2}
M^2_H - M^2_A\ \sim \ \pm\, 2\, \Im m\widehat{\Pi}^{HA}\, .
\end{equation}
Again, we remark that large mass differences, $M_H-M_A$, lead to small
CP-violating effects through particle mixing, {\em i.e.},
$|q/p|\approx 1$, near the region of the $H$ and $A$ resonances,
{\em e.g.}, for $s\approx M^2_H$ or $s\approx M^2_A$.

\end{itemize}

It is important to note that for maximal CP violation, {\em i.e.}, of
order unity, the value of $|q/p|$ should either vanish or tend to
infinity. This implies that either $\tilde{{\cal H}}_{12}=0$ or
$\tilde{{\cal H}}_{21}=0$, but {\em not } both. As will be discussed
in Section \ref{sec:WW}, this limiting case reflects the fact that the two
(non-free) particles, $H$ and $A$, are exactly degenerate, {\em i.e.},
$\overline{M}_H = \overline{M}_A$ and $\overline{\Gamma}_H =
\overline{\Gamma}_A$, where $\overline{M}^2_{H,A} -
i\overline{M}_{H,A}\overline{\Gamma}_{H,A}$ are the two complex
pole-mass eigenvalues of the effective Hamiltonian $\tilde{{\cal
H}}(s)$ (${\cal H}(s)$) in Eq.\ (\ref{KKHam}) (Eq.\ (\ref{InvDHA})).

Although our derivation of the necessary conditions for resonant CP
violation through particle mixing has been performed in a
$K^0\bar{K}^0$-like basis, we should stress again that these results
can apply equally well to any other orthogonal weak basis. In general,
the necessary conditions for CP invariance within perturbation field
theories may be expressed in terms of CP-odd invariants, which are
flavour-basis independent.  For example, there is only one such CP-odd
invariant in the minimal SM \cite{CJ,BBG}. In particular, the authors
in \cite{BBG} have devised a systematic approach to constructing
CP-odd invariants, by making use of generalized CP transformations.
Under generalized CP transformations, the fields are mapped into their
CP conjugates and, on the same footing, their weak basis is changed by
an orthogonal or unitary rotation, so as to leave the pure gauge
sector of the underlying Lagrangian invariant.  This approach has
found numerous applications to many extensions of the SM, including
those to multi-Higgs doublet models \cite{Lav/Sil}.  In Ref.\ 
\cite{Lav/Sil}, it has been concluded that there is no mixing-induced
CP violation at the tree-level of the two-Higgs doublet models, if two
Higgs particles have equal masses, {\em e.g.}, $M_H=M_A$. This result
is valid in the mass basis, in which $\Re e \widehat{\Pi}^{HA} = 0$,
{\em i.e.}, for the case (ii) discussed above. Indeed, our conclusions
are consistent with this condition. As can be seen from Eqs.\ 
(\ref{q/p}) and (\ref{q/p2}), we find that $|q/p|\to 1$, in the
mass-degenerate limit $M^2_H\to M^2_A$, at the resonant region
$s\approx M^2_H\approx M^2_A$. However, studying CP-odd invariants
only is not sufficient to obtain definite theoretical predictions
about the magnitude of CP violation. According to the necessary
conditions (\ref{CPN1}) and (\ref{CPN2}), even a small mass splitting
between $H$ and $A$ comparable to their widths can produce large CP
violation. A phenomenological example of the kind will be analyzed in
Section \ref{sec:CPHA}.

In the above discussion of CP-odd invariants, it has been very crucial
to differentiate the OS-renormalized masses $M_H$ and $M_A$ present in
our Hermitian Lagrangian from the mass eigenvalues $\overline{M}_H$
and $\overline{M}_A$ of the non-Hermitian effective Hamiltonian
(\ref{KKHam}). The former correspond to states defined in an
orthogonal or unitary Hilbert space, upon which our perturbation field
theory is based, whereas the resulting eigenvectors of the latter are
in general non-unitary or non-orthogonal among themselves. Since the
afore-mentioned generalized CP transformations refer to orthogonal
and/or unitary states, depending on whether the fields are real or
complex, this property of orthogonality and/or unitarity must be
maintained for the renormalized fields as well. Therefore, the CP-odd
invariants derived with the method in \cite{BBG} can only involve
masses that are renormalized within a well-defined field-theoretic
framework, which respects the property of unitarity, such as the
OS renormalization scheme \cite{Ren}.  The relation between OS and
pole-mass renormalization in the presence of a large particle mixing
is discussed in Appendix B. Moreover, CP-odd invariants have been
barely investigated beyond the Born approximation. In fact, not only
the masses of particles but also the mixing matrices must be
renormalized \cite{Sirlin1,BK&AP}.  Within the context of the approach
in \cite{BBG}, it is necessary to have a renormalization scheme that
consistently preserves the orthogonality or the unitarity \cite{BK&AP}
of the weak-basis transformations order by order in perturbation
theory. In Appendix A, we discuss how mixing-matrix renormalization
applies to scalar theories at one loop.

\setcounter{equation}{0}
\section{\label{sec:CPT} Constraints from CPT invariance}

So far, we have not taken into account constraints on transition
amplitudes that may arise from CPT invariance. The fact that the
underlying Lagrangian of the theory is CPT invariant gives rise to
additional relations between kinematic parameters and interactions.
These relations can, in principle, affect the size of the CP-violating
mixing phenomenon mentioned in Section \ref{sec:RCPV}. In general,
they have the tendency to reduce the actual magnitude of $a_{CP}$ in
Eq.\ (\ref{aCP}), even though $|q/p|$ takes its extreme value. In
this section, we shall briefly illustrate how CPT restoration takes
place.

Consider the transition of some initial state $i$ to a final state
$f$.  Then, up to an insignificant overall phase, CPT invariance
requires
\begin{equation}
\label{CPT}
\langle f|T|i\rangle \ =\ \langle i_{CPT}|T|f_{CPT}\rangle\, ,
\end{equation}
where the subscript CPT indicates that the states $i$ and $f$ are
transformed under CPT. For some state $\beta$ with three-momentum
$\vec{p}$ and spin $\vec{s}$, CPT acts as
$|\beta_{CPT}(\vec{p},\vec{s})\rangle = |\bar{\beta}
(\vec{p},-\vec{s})\rangle$, with $\bar{\beta}$ being the anti-particle
of $\beta$. Equation (\ref{CPT}) holds for the Hermitian and
anti-Hermitian part of ${\cal T}_{fi}\equiv \langle f|T|i\rangle$
independently. As a consequence of (\ref{CPT}), using the optical
theorem,
\begin{equation}
\label{OT}
{\cal T}_{fi}\, -\, {\cal T}^*_{if}\ =\ i\sum\limits_k\, {\cal
T}_{ki}{\cal T}_{kf}^*\, ,
\end{equation}
where the sum is over all possible intermediate states $k$ including
phase-space integration, and taking $|i\rangle = |f\rangle \equiv
|\alpha\rangle$, one obtains the relation
\begin{equation}
\label{CPTwid}
\Im m\langle \alpha |T| \alpha\rangle\ =\ \Im m\langle \alpha_{CPT}
|T| \alpha_{CPT}\rangle\ =\ m_\alpha \sum\limits_{X}\, \Gamma (\alpha
\to X)\, ,
\end{equation} 
where $m_\alpha$ is the mass of the decaying particle $\alpha$ and the
summation is understood over all possible final states $X$ that
$\alpha$ can decay. Equation (\ref{CPTwid}) together with the fact
that the Hamiltonian of the theory is Hermitian represent the known
corrolary emanating from CPT invariance, which states that the mass
and lifetime of a particle is equal with that of its anti-particle.

It is now straightforward to extend our considerations to scattering
processes. Analogously with Eq.\ (\ref{CPTwid}), taking as asymptotic
states $|i\rangle = |f\rangle = |a (\vec{p}_a,\vec{s}_a) b
((\vec{p}_b,\vec{s}_b) \rangle$, and making use of the optical theorem
in Eq.\ (\ref{OT}) for the forward scattering, {\em viz.}
\begin{equation}
\label{OTforw}
\Im m \langle a (\vec{p}_a,\vec{s}_a) b (\vec{p}_b,\vec{s}_b)|\, T\,
|a (\vec{p}_a,\vec{s}_a) b (\vec{p}_b,\vec{s}_b) \rangle\, =\,
\lambda^{1/2}(s,m^2_a,m^2_b)\, \sum\limits_X \sigma \big( a
(\vec{p}_a,\vec{s}_a) b (\vec{p}_b,\vec{s}_b) \to X \big),
\end{equation}
and for its CPT-conjugate counterpart, with $\lambda (x,y,z)=
(x-y-z)^2-4yz$, we arrive at the CPT constraint involving total cross
sections
\begin{equation}
\label{CPTcros}
\sum\limits_X \sigma \big( a (\vec{p}_a,\vec{s}_a) b
(\vec{p}_b,\vec{s}_b) \to X \big)\ =\ \sum\limits_X \sigma \big(
\bar{a} (\vec{p}_a,-\vec{s}_a) \bar{b} (\vec{p}_b,-\vec{s}_b) \to X
\big)\, .
\end{equation}
Note that Eqs.\ (\ref{CPTwid}) and (\ref{CPTcros}) are still valid
within the perturbation theory through the order considered.

\begin{center}
\begin{picture}(440,100)(0,0)
\SetWidth{0.8}

\Vertex(0,50){1.7} \Line(0,50)(35,50)\Text(0,62)[l]{$K^0$}
\Line(35,50)(70,50)\Text(60,62)[]{$K^0,\bar{K}^0$}
\GCirc(35,50){10}{0.5} \Line(70,50)(90,70) \Line(70,50)(90,30)

\Text(100,20)[]{$X$}

\Line(110,30)(130,50) \Line(110,70)(130,50)
\Line(130,50)(165,50)\Text(145,62)[]{$K^0,\bar{K}^0$}
\Line(165,50)(200,50)\Text(185,62)[]{$K^0$} \GCirc(165,50){10}{0.5}
\Vertex(200,50){1.7}

\Text(100,0)[]{{\bf (a)}}


\Vertex(240,50){1.7} \Line(240,50)(275,50)\Text(255,62)[]{$\bar{K}^0$}
\Line(275,50)(310,50)\Text(300,62)[]{$K^0,\bar{K}^0$}
\GCirc(275,50){10}{0.5} \Line(310,50)(330,70) \Line(310,50)(330,30)

\Text(340,20)[]{$X$}

\Line(350,70)(370,50) \Line(350,30)(370,50)
\Line(370,50)(405,50)\Text(385,62)[]{$K^0,\bar{K}^0$}
\Line(405,50)(440,50)\Text(430,62)[]{$\bar{K}^0$}
\GCirc(405,50){10}{0.5} \Vertex(440,50){1.7}

\Text(340,0)[]{{\bf (b)}}

\end{picture}\\[0.7cm]
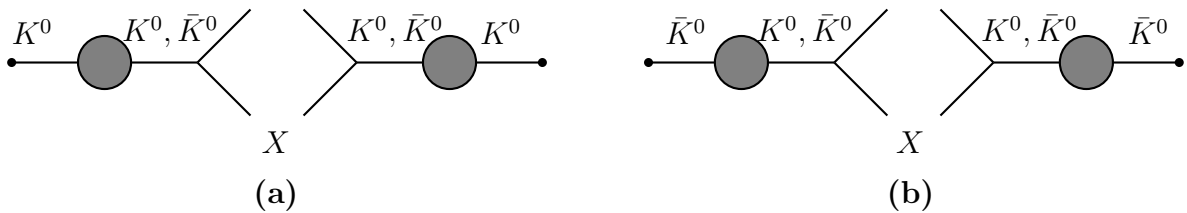
{\small {\bf Fig.\ 3:} CPT invariance in the squared transition
amplitudes.}
\end{center}

It is now interesting to see how CPT applies to our resummation
formalism.  To facilitate our task, we assume that there are some
asymptotic states, $a$ and $b$ say, that couple to $K^0$ only and
produce it with an effective vertex $V_K$. Then, the CPT-conjugate
states, $a_{CPT}$ and $b_{CPT}$, will only couple to $\bar{K}^0$ with
a production amplitude $\bar{V}_K$. We also assume that $a$ and $b$ do
not introduce $\varepsilon'$ effects into the vertex
$V_K$. Subsequently, the so-produced kaons, $K^0$ and $\bar{K}^0$,
decay into all possible final states $X$ with couplings $V^X_1$ and
$V^X_2$, respectively, as shown in Fig.\ 3. Taking only the dominant
$s$-channel contributions into account, the two squared amplitudes can
be cast into the form
\begin{eqnarray}
\label{TK}
|{\cal T}_K|^2 & =& V_K\, \big( s-\tilde{{\cal H}}\big)^{-1}_{1i}\,
V^X_i\, V^{X*}_j\, \big( s-\tilde{{\cal H}}^\dagger\big)^{-1}_{j1}\,
V^*_K\, ,\\
\label{TKbar}
|\overline{{\cal T}}_K|^2 & =& \bar{V}_K\, \big( s-\tilde{{\cal
        H}}\big)^{-1}_{2i}\, V^X_i\, V^{X*}_j\, \big( s-\tilde{{\cal
        H}}^\dagger\big)^{-1}_{j2}\, \bar{V}^*_K\, ,
\end{eqnarray}
where summation over the repeated indices $i,j=1,2$ and integration
over the phase space of the final states $X$ must be understood.
Employing the optical theorem in Eq.\ (\ref{OT}) at the level of the
effective Hamiltonian $\tilde{{\cal H}}$, one has through the order
considered
\begin{equation}
\label{OTH}
\big(\, \tilde{{\cal H}}\, -\, \tilde{{\cal H}}^\dagger\, \big)_{ij}\
=\ -\, i\, V^X_i V^{X*}_j.
\end{equation}
Substituting Eq.\ (\ref{OTH}) into Eqs.\ (\ref{TK}) and (\ref{TKbar}),
it is not difficult to find
\begin{equation}
\label{CPTKK}
|{\cal T}_K|^2\ =\ -2\, |V_K|^2 \Im m \big( s-\tilde{{\cal
H}}\big)^{-1}_{11}\, , \qquad |\overline{{\cal T}}_K|^2\ =\ -2\,
|\bar{V}_K|^2 \Im m \big( s-\tilde{{\cal H}}\big)^{-1}_{22}\, .
\end{equation}
As has been assumed above, absence of CP violation in the production
vertices implies that $|V_K|=|\bar{V}_K|$. Since the effective
Hamiltonian is CPT invariant, {\em i.e.}, $\tilde{{\cal
H}}_{11}=\tilde{{\cal H}}_{22}$ ({\em cf.}\ Eq.\ (\ref{CPTHam})), one
easily concludes that $|{\cal T}_K|^2=|\overline{{\cal T}}_K|^2$, as
it should be in agreement with Eq.\ (\ref{CPTcros}).

It is evident that CPT symmetry yields relations among the
CP-violating parts of the squared transition amplitudes similar to
those found for the partial decay rates in Ref.\ \cite{LW}. To make
this explicit, let us consider the example mentioned above and assume
that the set of all final states $X$ can be divided into two
sub-sets. The first sub-set involves the states $a,b$, {\em i.e.},
$a,b\to K^{0*} \to a,b$, while the second one, $A$ say, does not
contain the states $a,b$.  Thus, the squared amplitudes governing the
reaction $a,b\to K^{0*} \to A$ and its CP-conjugate process are given
by
\begin{eqnarray}
\label{TKA}
|{\cal T}_{AK}|^2 &=& |V_K|^2\, (s-\tilde{{\cal H}})^{-1}_{1i} V^A_i
V^{A*}_j (s-\tilde{{\cal H}}^\dagger )^{-1}_{j1}\, , \\
\label{TCPKA}
|{\cal T}_{\bar{A}\bar{K}}|^2 &=& |\bar{V}_K|^2\, (s-\tilde{{\cal
    H}})^{-1}_{2i} \bar{V}^A_i\bar{V}^{A*}_j (s-\tilde{{\cal
    H}}^\dagger )^{-1}_{j2} \nonumber\\ &=& |V_K|^2\, (s-\tilde{{\cal
    H}}^T)^{-1}_{1i} V^A_i V^{A*}_j (s-\tilde{{\cal H}}^*
    )^{-1}_{j1}\, .
\end{eqnarray}
In the last equality of Eq.\ (\ref{TCPKA}), we have assumed that CP
violation is completely absent in the production and decay vertices,
{\em i.e.}, $V_K=\bar{V}_K$, $V^A_1=\bar{V}^A_2$ and
$V^A_2=\bar{V}^A_1$, which amounts to $\Im m\Pi^{HA}=0$ in the $HA$
basis. In the subsequent sections, we shall see that expressions
analogous to Eqs.\ (\ref{TKA}) and (\ref{TCPKA}) will be used in
explicit calculations of CP violation.  For this purpose, we introduce
the following difference based on matrix elements squared:
\begin{equation}
\label{DAK}
\Delta_{AK}\ =\ |{\cal T}_{AK}|^2\, -\, |{\cal
T}_{\bar{A}\bar{K}}|^2\, ,
\end{equation}
which is a genuine CP-violating quantity. Similarly, one can define the
CP-violating difference $\Delta_{KK}$ as 
\begin{equation}
\label{DKK}
\Delta_{KK}\ =\ |{\cal T}_{KK}|^2\, -\, |{\cal
T}_{\bar{K}\bar{K}}|^2\, ,
\end{equation}
for the process $a,b\to K^{0*} \to a,b$ and its CP-conjugate reaction.
As a consequence of CPT invariance in Eq.\ (\ref{CPTKK}), one has that
\begin{equation}
\label{CPTDs}
\Delta_{KK}\ +\ \Delta_{KA}\ =\ 0.
\end{equation}
The equality (\ref{CPTDs}) demonstrates explicitly how the sum of all
partial CP-violating differences vanishes \cite{LW} within this
formalism.

Another important consequence of CPT symmetry is the decrease of CP
violation, as observed by $a_{CP}$ in Eq.\ (\ref{aCP}), despite the
fact that the CP-asymmetric mixing as expressed by the parameter
$|q/p|$ is maximal. To exemplify this point, we consider the effective
Hamiltonian
\begin{equation}
\label{Hextr}
\tilde{{\cal H}}\ =\ \left[\begin{array}{cc} M-i\Gamma &-2i\Gamma_{12}
\\ 0 & M-i\Gamma \end{array} \right]\, ,
\end{equation}
which corresponds to the extreme value of $|q/p|=0$. Clearly, the
entries of the effective Hamiltonian in Eq.(\ref{Hextr}) are given in
units of energy squared. Defining as
\begin{eqnarray}
\label{VAs}
iV^A_1 V^{A*}_1 & =& iV^A_2V^{A*}_2\ =\ 2i\bar{\Gamma}\ =\ 2i\Gamma -
i|V_K|^2, \nonumber\\ iV^A_1 V^{A*}_2 & =& iV^A_2V^{A*}_1\ =\
2i\bar{\Gamma}_{12}\ =\ 2i\Gamma_{12} - iV_K\bar{V}^*_K\, ,
\end{eqnarray}
we find at energies $s\approx M$ that for the reaction $a,b\to
K^{0*}\to A$, $a_{CP}$ takes the simple form
\begin{equation}
\label{aCP1}
a_{CP}\ =\ \frac{\Delta_{AK}}{|{\cal T}_{AK}|^2\, +\, |{\cal
T}_{\bar{A}\bar{K}}|^2}\ =\ \frac{2\Gamma_{12}
(\bar{\Gamma}\Gamma_{12}\, -\, \Gamma\bar{\Gamma}_{12})}{
\Gamma^2\bar{\Gamma}\, +\, 2\Gamma_{12} (\bar{\Gamma}\Gamma_{12}\, -\,
\Gamma\bar{\Gamma}_{12})}\, .
\end{equation}
From Eq.\ (\ref{aCP1}), one can readily see that $a_{CP}$ vanishes if
$A$ represents the sum over all final states $X$, as a result of CPT.
Furthermore, it is obvious that the value of $a_{CP}$ is generally
smaller than unity even in the extreme particle-mixing limit. In
fact, $a_{CP}$ can take its maximum value only in the limit
$\bar{\Gamma}\to 0$, with $\bar{\Gamma}_{12}\not= 0$. In such a case,
the process $\bar{a}\bar{b}\to \bar{K}^{0*}\to \bar{A}$ is forbidden,
whereas the transition $ab\to K^{0*}\to A$ is allowed. Hence, we can
conclude that our resummation formalism respects the CPT symmetry and
takes automatically account of all possible CPT constraints.

\setcounter{equation}{0}
\section{\label{sec:WW} Effective Hamiltonian approach}

We shall briefly review the main features of the effective Hamiltonian
approach, which was formulated by Weisskopf and Wigner \cite{WW} and
applied to describe CP violation in the kaon system by Lee, Oehme and
Yang \cite{LOY}. Our approach presented in Sections \ref{sec:Bos} and
\ref{sec:RCPV} is equivalent with that based on diagonalizing the
effective Hamiltonian \cite{WW,LS}. In this section, we shall pay
special attention to the limitations of the latter that may arise from
describing CP violation in a system where the two mixed particles
become degenerate. In fact, we will show that there exist extreme
cases, for which the effective Hamiltonian cannot be diagonalized via
a similarity transformation.  Such non-diagonalizable effective
Hamiltonians represent situations, in which the mixing-induced CP
violation reaches its maximum allowed value and the complex mass
eigenvalues of the effective Hamiltonian are exactly degenerate. In
addition, we shall show that non-diagonalizable effective Hamiltonians
are admissible forms, for which the Lee-Wolfenstein \cite{Lee/Wol}
inequality due to unitarity is saturated.  In this context, we find that our
resummation approach based on transition amplitudes constitutes a
self-consistent field-theoretic framework for dealing even with such
singular cases.

To elucidate our points, we shall consider the $HA$ system from the
viewpoint of the effective Hamiltonian formalism. In the WW
approximation, the time evolution of the $HA$ system can be described
by means of an effective Schr\"odinger equation, in which the Hamilton
operator is a $2\times 2$ non-Hermitian matrix \cite{LOY}. Taking
dispersive and absorptive quantum effects into account, this effective
Hamiltonian is given by
\begin{equation}
\label{WW_H}
{\cal H}\ =\ \left[ \begin{array}{cc}
H_{11} & H_{12} \\ H_{21} & H_{22} \end{array} \right]\ =
\ \left[ \begin{array}{cc}
M_{11}-i\Gamma_{11} & M_{12}-i\Gamma_{12} \\ 
M^*_{12}-i\Gamma^*_{12} & M_{22}-i\Gamma_{22} \end{array} \right]\, ,
\end{equation}
where the index 1 (2) refers to $A$ ($H$) and all entries of ${\cal
  H}$ are given in units of energy squared.  As we have seen in
Section 2, the effective Hamiltonian in Eq.\ (\ref{WW_H}) may be
approximated by the matrix ${\cal H}(s)$ defined in Eq.\ (\ref{InvDHA}), for
$s\approx M^2_H\approx M^2_A$.  In the last equality of Eq.\ (\ref{WW_H}),
we have decomposed the non-Hermitian Hamiltonian in terms of two
Hermitian matrices as follows:
\begin{equation}
M_{ij}\ =\ \frac{1}{2}\, \big(\, {\cal H}\, +\, {\cal H}^\dagger\, \big)\, ,
\quad
\Gamma_{ij}\ =\ \frac{i}{2}\, \big(\, {\cal H}\, -\, {\cal H}^\dagger\, 
\big)\, .
\end{equation}
Furthermore, CPT invariance requires that $|H_{12}|=|H_{21}|$ for the
case of neutral-scalar mixing, which in turn implies that $\Im m
(M_{12}\Gamma^*_{12})=0$.  Since $M_{12}$ is real as being the
dispersive part of transitions among real scalars (see also Appendix
A), $\Gamma_{12}$ should be a real function as well, as a consequence
of CPT invariance.  However, we must caution the reader that this CPT
constraint does not apply for mixing between charged scalars.

Assuming the existence of a non-unitary matrix $X$, ${\cal H}$ can be
brought into a diagonal form through a similarity transformation
\begin{equation}
\label{Xtrafo}
X\, {\cal H}\, X^{-1}\ =\ \left( \begin{array}{cc}
\overline{M}^2_A\, -\ i\overline{\Gamma}_A\overline{M}_A & 0\\
0 & \overline{M}^2_H\, -\ i\overline{\Gamma}_H\overline{M}_H
\end{array} \right)\, ,
\end{equation}
where the two complex mass eigenvalues (or the two pole masses of the
transition amplitude \cite{Sachs}) are given by
\begin{eqnarray}
\label{polmas}
\overline{M}^2_A\, -\ i\overline{\Gamma}_A\overline{M}_A &
= & \frac{1}{2} \big( H_{11}\, +\, H_{22}\, -\, \Delta\, \big)\, ,\nonumber\\
\overline{M}^2_H\, -\ i\overline{\Gamma}_H\overline{M}_H &
= & \frac{1}{2} \big( H_{11}\, +\, H_{22}\, +\, \Delta\, \big)\, ,
\end{eqnarray}
where $\Delta^2\, =\, (H_{11}-H_{22})^2\, +\, 4H_{21}H_{12}$ and the
square root is taken on the first sheet. Clearly, the masses
$\overline{M}_H$ and $\overline{M}_A$ differ from those that the $H$
and $A$ particles would have if they were free, {\em i.e.}, if all
couplings were switched off. The matrices $X$ and $X^{-1}$ that
diagonalize ${\cal H}$ have the explicit forms:
\begin{eqnarray}
\label{Xmat}
X &=& \Big(  \frac{\Delta - H_{11} + H_{22}}{2 \Delta}\Big)^{1/2}\, 
\left[ \begin{array}{cc}
1&-\, \frac{\displaystyle 1}{\displaystyle 2H_{21}}(H_{11}-H_{22}+\Delta)\\
\frac{\displaystyle 1}{\displaystyle 2H_{12}}(H_{11}-H_{22}+\Delta)&1
\end{array} \right]\, ,\qquad\\
\label{Xmat1}
X^{-1} &=& \Big(  \frac{\Delta - H_{11} + H_{22}}{2 \Delta }\Big)^{1/2}\, 
\left[ \begin{array}{cc}
1&\frac{\displaystyle 1}{\displaystyle 2H_{21}}(H_{11}-H_{22}+\Delta)\\
-\, \frac{\displaystyle 1}{\displaystyle 2H_{12}}(H_{11}-H_{22}+\Delta)&1
\end{array} \right]\, .\qquad
\end{eqnarray}
From Eqs.\ (\ref{Xmat}) and (\ref{Xmat1}), one may easily deduce that
the matrices $X$ or $X^{-1}$ do exist unless both $\Delta = 0$ and the 
off-diagonal elements $H_{12}$, $H_{21}$ are different from zero. 
Of course, if $H_{12}=H_{21}=0$, $X$ becomes the unity matrix. However,
one could think of an effective Hamiltonian with $\Delta =0$ and 
$H_{12}=H_{21}\not= 0$, for which $X$ becomes {\em singular}, {\em e.g.}
\begin{equation}
\label{Hanom}
{\cal H} \ =\ \left[ \begin{array}{cc}
A-ib & -b/2 \\
-b/2 & A-2ib \end{array} \right]\, ,
\end{equation}
where $A$ and $b$ are some positive numbers. Most importantly, the two
pole-mass  eigenvalues, for $H$ and $A$  say, are {\em exactly} equal,
since  $\Delta =0$ in Eq.\  (\ref{polmas}).  As a consequence, the two
{\em non-free} particles $H$ and $A$ are {\em exactly} degenerate.  As
has been found in \cite{APRL} and will be  further analyzed in Section
\ref{sec:CPHA},  such a singular case can   occur in two-Higgs doublet
models, when one studies resonant  CP violation due  to a $HA$ mixing.
For example, if the Higgs self-energies in Eq.\ (\ref{InvDHA}) satisfy
the  relation:  $b  = \Im  m   \widehat{\Pi}^{AA} = \frac{1}{2} \Im  m
\widehat{\Pi}^{HH} =  2\widehat{\Pi}^{HA}$ for  $s =  M^2_H  = M^2_A$,
this gives rise to the non-diagonalizable effective Hamiltonian of the
form (\ref{Hanom}).

Apart from the difficulty of diagonalizing ${\cal H}$ for a degenerate
system of bound states, one may now raise the following question: What is 
the actual magnitude of CP violation through particle mixing in such a 
degenerate case? To successfully address this question, one should 
perform the above analysis in a $K^0\bar{K}^0$-like basis. Taking
the basis rotations in Eq.\ (\ref{KKtrafo}) into account, we 
obtain the new effective Hamiltonian $\tilde{{\cal H}}$ (equivalent
to Eq.\ (\ref{KKHam})) and the respective non-unitary matrix $\tilde{X}$
\begin{equation}
\tilde{{\cal H}}\ =\ \frac{1}{2}\, \left[ \begin{array}{cc}
H_{11}+H_{22} & H_{22}- H_{11} + 2iH_{12} \\
H_{22}- H_{11} - 2iH_{12} & H_{11}+H_{22} \end{array} \right]\, ,
\qquad \tilde{X}\ =\ \frac{1}{\sqrt{2}}\left[ \begin{array}{cc}
1 & -p/q \\ q/p & 1 \end{array} \right]\, ,
\end{equation}
where $q/p$ is the basic parameter, which determines the CP-violating
phenomenon induced by the CP-asymmetric mixing of $K^0$ with its
CP-conjugate state, $\bar{K}^0$ \cite{review}.  This parameter 
is given by
\begin{equation}
\label{qp}
\frac{q}{p}\ =\ \Big( \frac{\tilde{{\cal H}}_{21}}{\tilde{{\cal 
H}}_{12}}\Big)^{1/2}.
\end{equation}
In the effective Hamiltonian formalism, the two (right) mass eigenstates, 
$|K_L\rangle$ and $|K_S\rangle$, are then expressed in terms of the strong 
states, $K^0$ and $\bar{K}^0$, as\footnote[1]{Eq.\ (\ref{KLKS}) differs
from the standard convention, since we have defined CP$(K^0)=\bar{K}^0$
instead of $-\bar{K}^0$ and assigned the two pole-mass eigenstates
$K_L$ and $K_S$ in the reversed order.}
\begin{equation}
\label{KLKS}
|K_L\rangle\, =\, q|K^0\rangle\, -\, p|\bar{K}^0\rangle\quad
\mbox{and}\quad
|K_S\rangle\, =\, q|K^0\rangle\, +\, p|\bar{K}^0\rangle\, ,\quad
\end{equation}
with the normalization $|p|^2+|q|^2=1$.  In this phase convention, the
CP-violating mixing parameter $\varepsilon =\langle K_L | K_S \rangle$
is a real number.  Obviously, the anomalously degenerate $HA$ or
$K^0\bar{K}^0$ system mentioned above, which leads to a singular
$\tilde{X}$, corresponds to the case where either $\tilde{{\cal
    H}}_{21}$ or $\tilde{{\cal H}}_{12}$ vanish but {\em not} both of
them, {\em i.e.}, the effective Hamiltonian $\tilde{\cal H}$ takes
mathematically the Jordan form \cite{Jordan}. In such a case, $q/p$ is
either zero or tends to infinity, which shows explicitly that CP
violation through particle mixing takes its maximum attainable value.
Apart from CPT constraints, CP violation could be of order unity, when
the non-free $H$ and $A$ states are exactly degenerate. 

It   is now important  to  investigate  whether the non-diagonalizable
Jordan  form of the  effective Hamiltonian  satisfies unitarity. Using
the optical   theorem  in  Eq.\  (\ref{OT})  for  a  general effective
Hamiltonian    $\tilde{{\cal  H}}$,    one   can   easily  obtain  the
Bell-Steinberger unitarity relation \cite{Bel/Stein}
\begin{eqnarray}
\label{BSrel}
\langle K_L | (\tilde{\cal H}\, -\, \tilde{\cal H}^\dagger) | K_S \rangle &=&
[\, \overline{M}^2_H -\overline{M}^2_A - i(\overline{M}_A\overline{\Gamma}_A
+\overline{M}_H\overline{\Gamma}_H)]\, \langle K_L| K_S \rangle
\nonumber\\
&=& i\, \sum_\xi \langle \xi |T|K_L\rangle^* \langle \xi |T|
K_S\rangle\, ,
\end{eqnarray}
where $\xi$ represents all intermediate states that $K_L$ and $K_S$
can decay. In this notation, $K_L$ and $K_S$ formally denote the two non-free
$A$ and $H$ mass-eigenstates of $\tilde{{\cal H}}$, respectively. 
Employing Schwartz's inequality 
\begin{equation}
\label{Swartz}
\Big|\sum_\xi \langle \xi |T|K_L\rangle^* \langle \xi |T|
K_S\rangle\Big|^2\ \leq \ \Big|\sum_m \langle m |T|K_L\rangle\Big|^2\ 
\Big|\sum_n \langle n |T|K_S\rangle\Big|^2\, ,
\end{equation}
and the fact that $\sum_\xi
|\langle \xi | T | X\rangle |^2 = 2\overline{M}_X\overline{\Gamma}_X$,
for $X=K_L,\ K_S$, we find the following inequality:
\begin{equation}
\label{LWrel}
|\varepsilon |^2\ =\ |\langle K_L| K_S\rangle|^2\ \leq \ 
\frac{4\overline{M}_H\overline{\Gamma}_H
\overline{M}_A\overline{\Gamma}_A}{(\overline{M}_H^2
-\overline{M}_A^2)^2\, +\, (\overline{M}_H\overline{\Gamma}_H+
\overline{M}_A\overline{\Gamma}_A)^2 }\ ,
\end{equation}
which is known as the Lee-Wolfenstein  inequality \cite{Lee/Wol}.  The
relation between the CP-violating  mixing parameters $\varepsilon$ and
$|q/p|$ is given by $|q/p|^2 = (1 + \varepsilon )/(1 - \varepsilon )$.
Clearly, the  non-diagonalizable    Jordan form  of   $\tilde{\cal H}$
corresponds to the  values of $\varepsilon =  \pm 1$.  For the extreme
case   of    exact mass degeneracy,  {\em     i.e.}, $\overline{M}_H =
\overline{M}_A$ and  $\overline{\Gamma}_H  = \overline{\Gamma}_A$,  we
obtain  the  unitarity  restriction $|\varepsilon|^2   \leq  1$, as is
derived  from    Eq.\  (\ref{LWrel}).   It    is  obvious   that   the
Lee-Wolfenstein   unitarity    bound     gets   saturated     in   the
non-diagonalizable  limit  of  the  effective Hamiltonian $\tilde{\cal
  H}$.

Our   resummation approach is    free from  the kind  of  pathological
singularities mentioned above, which  arise  whenever one attempts  to
diagonalize the   effective  Hamiltonian \cite{LOY}   or  the  inverse
propagator $\hat{\Delta}^{-1} (s)$ in Eq.\ (\ref{InvDHA}) \cite{Sachs}
for an anomalously degenerate system.   Since we always consider  {\em
  physical} transition  amplitudes, in  which the resummed propagators
are {\em sandwiched} between    the  matrix elements related   to  the
initial  and final states of the  process,  the diagonalization of the
propagator  $\hat{\Delta} (s)$ is {\em  no more} necessary.  Thus, our
scattering-amplitude  approach   is reminiscent of the   known density
matrix formalism \cite{DMF}.  In  fact, as the two  non-free particles
approach the  anomalously   degenerate limit,  the  mass  eigenvectors
obtained from ${\cal H}$ or  $\tilde{{\cal H}}$, such as $|K_L\rangle$
and    $|K_S\rangle$  as  well   as  their   dual  (left) eigenvectors
\cite{Sachs,LStod}, may  not bear  any physical meaning \cite{Jordan}.
Only  the  complex pole positions  of  a  transition amplitude  can be
considered  as physical (observable)  quantities.  Of course, far away
from  the  anomalous   situation,   the traditional    description  of
diagonalizing  the effective Hamiltonian  \cite{LOY}  can equally well
represent  the known  experimental  data, namely,  CP violation in the
ordinary $K^0\bar{K}^0$ system \cite{review}.

\setcounter{equation}{0}
\section{\label{sec:HA} {\boldmath $HA$} mixing in two-Higgs doublet models}

New-physics scenarios that could predict a potentially large $HA$
mixing should have two ingredients: First, they should extend the
field content of the SM by an extra Higgs doublet and, second, they
must violate the CP symmetry of the Higgs sector. There are two
representative paths that can lead to a CP-violating $HA$ term. The
first possibility is to produce such a term at the tree-level. We
shall see that this purpose can naturally be achieved within a two-Higgs
doublet model, in which the CP invariance of the Higgs potential is
explicitly broken by adding soft mass terms. Another alternative is to
induce a $HA$ mixing radiatively after integrating out heavy degrees
of freedom.  These heavy degrees of freedom will explicitly violate
the CP symmetry of the Lagrangian. As such, one may think of heavy
Majorana fermions, such as heavy Majorana neutrinos or neutralinos in
the minimal supersymmetric SM (MSSM), that couple to $H$ and $A$
scalars with both scalar and pseudo-scalar couplings. In the
following, we shall determine the size of the CP-violating $HA$ term
in the two models mentioned above.

\noindent (i) {\em $HA$ mixing in the two-Higgs doublet model with an
explicitly CP-violating Higgs potential}. First, we shall briefly
describe the basic structure of this minimally extended model, which
may be considered as the simplest realization of CP violation in the
neutral Higgs sector.  The presence of two-Higgs doublets, $\Phi_1$
and $\Phi_2$, in the Lagrangian can give rise to large
flavour-changing neutral currents (FCNC) in the Higgs coupling to
fermions. In order to avoid FCNC at the tree level, one usually
imposes the discrete symmetry D: $\Phi_1\to -\Phi_1$, $\Phi_2\to
\Phi_2$ and $d'_{iR}\to -d'_{iR}$, where $d'_{iR}$ are the
right-handed down-type quarks in the flavour basis. Invariance under
the discrete symmetry D entails that $\Phi_1$ will couple to down-type
quarks and $\Phi_2$ to up-type quarks only, thus leading to diagonal
couplings of the neutral Higgs particles to quarks
\cite{BSP,PN}. Furthermore, imposing the D symmetry on the Higgs
potential leads to a CP-invariant two-Higgs doublet model. The model
predicts four neutral scalar fields: the three massive fields $h$,
$H$, $A$ and the massless would-be Goldstone boson of the $Z$ boson,
$G^0$. In the D-symmetric limit, the neutral Higgs scalars $h$ and $H$
are exactly CP even, whereas the third neutral Higgs particle, $A$, is
CP odd. The Lagrangians describing the interactions of the neutral
Higgs scalars to quarks are given by
\begin{eqnarray}
\label{LG0ff}
{\cal L}_{G^0} &=& \frac{igm_f}{2M_W}\, G^0\, \bar{f} T^f_z\gamma_5
f\, ,\\
\label{LAff}
{\cal L}_A &=& \frac{ig}{2M_W}\, A\, \Big[\, \cot\beta\, m_{u_i}\,
\bar{u}_i \gamma_5 u_i\ +\ \tan\beta\, m_{d_i}\, \bar{d}_i \gamma_5
d_i\, \Big]\, ,\\
\label{LHhff}
{\cal L}_{H,h} &=& -\frac{g}{2M_W}\, \Big[\, (h \chi^u_h\, +\,
H\chi^u_H )\, m_{u_i}\, \bar{u}_i u_i\, +\, (h\chi^d_h\, +\, H\chi^d_H
)\, m_{d_i}\, \bar{d}_i d_i\, \Big]\, ,\quad
\end{eqnarray}
with
\begin{equation}
\label{chis}
\chi^u_h\, =\, -\frac{\sin\theta}{\sin\beta}\, ,\quad \chi^u_H\, =\,
\frac{\cos\theta}{\sin\beta}\, ,\quad \chi^d_h\, =\,
\frac{\cos\theta}{\cos\beta}\, ,\quad \chi^d_H\, =\,
\frac{\sin\theta}{\cos\beta}\, .
\end{equation}
Here, $g$ is the weak coupling constant, $\tan\beta = v_2/v_1$ is the
ratio of the $\Phi_2$ to $\Phi_1$ vacuum expectation values (VEV's),
the angle $\theta$ relates the weak states to the physical states $h$
and $H$, and $T^f_z$ is the $z$-component of the weak isospin of the
fermion $f$ ($T^u_z=1$, $T^d_z=-1$).  The absence of FCNC in the Higgs
couplings can easily be extended to the lepton sector, by requiring in
addition to symmetry D that $l'_{iR}\to -l'_{iR}$, where $l'_{iR}$ are
the right-handed charged leptons. The interaction Lagrangians for the
Higgs particles with the charged leptons $l_i = e$, $\mu$ and $\tau$ may
then be obtained from Eqs.\ (\ref{LG0ff})--(\ref{LHhff}), after making
the obvious replacements: $d\to l$ and $m_{d_i} \to m_{l_i}$.

Up to now, CP has been a good symmetry of the whole Lagrangian
provided the discrete symmetry D remains unbroken. In order to
introduce CP violation in the theory, we must break the symmetry D in
some part of the Lagrangian. The most convenient way is to write down
soft-breaking mass terms in the Higgs sector that violate the symmetry
D explicitly \cite{DM}.  In this way, the Higgs potential $V_H$ may be
decomposed into two terms as follows:
\begin{equation}
\label{VH}
V_H (\Phi_1,\Phi_2)\ =\ V_D(\Phi_1,\Phi_2)\, +\, \Delta
V(\Phi_1,\Phi_2)\, ,
\end{equation}
where $V_D$ respects the D symmetry and has the general Hermitian form
\begin{eqnarray}
\label{VD}
V_D (\Phi_1,\Phi_2 ) &=& \mu^2_1 (\Phi_1^\dagger\Phi_1)\, +\, \mu^2_2
(\Phi_2^\dagger\Phi_2)\, +\, \frac{1}{2}\, \lambda_1
(\Phi_1^\dagger\Phi_1)^2\, +\, \frac{1}{2}\, \lambda_2
(\Phi_2^\dagger\Phi_2)^2\nonumber\\ && +\, \lambda_3
(\Phi_1^\dagger\Phi_1) (\Phi_2^\dagger\Phi_2)\, +\, \lambda_4
(\Phi_1^\dagger\Phi_2) (\Phi_2^\dagger\Phi_1)\nonumber\\ &&+\,
\frac{1}{2}\, \lambda_5 (\Phi_1^\dagger\Phi_2)^2\, +\, \frac{1}{2}\,
\lambda^*_5 (\Phi_2^\dagger\Phi_1)^2\, ,
\end{eqnarray}
whereas $\Delta V$,
\begin{equation}
\label{DVD}
\Delta V (\Phi_1,\Phi_2)\ =\ \lambda_6 (\Phi^\dagger_1\Phi_2)\ +\
\lambda^*_6 (\Phi^\dagger_2\Phi_1)\, ,
\end{equation}
violates the symmetry D softly. Invariance of $V_H$ under CP is only
reassured if
\begin{equation}
\label{CPDVD}
\Im m (\lambda_5\lambda^{*2}_6)\ =\ 0\, .
\end{equation}
After SSB, the VEV's of $\Phi_1$ and $\Phi_2$, $v_1$ and $v_2$, are
generally complex relative to one another, {\em i.e.}, $\Im m
(v^*_1v_2) \not= 0$, if CP is not conserved. Nevertheless, using the
freedom of the phase redefinitions, $\Phi_i \to e^{i\alpha_i}\Phi_i$,
in $V_H$, we can always make $v_1$ and $v_2$ real at the cost of both
$\lambda_5$ and $\lambda_6$ being complex. As a result, any
CP-noninvariant term will be proportional either to $\Im m
(\lambda_5\lambda^{*2}_6)$ or $\Im m\lambda_5$. The latter can,
however, be expressed in terms of the former by making use of the
conditions obtained from minimizing the Higgs potential $V_H$. This
D-broken two-Higgs doublet model predicts CP-violating mass terms,
such as $HA$ and $hA$, already at the tree level. Specifically, we
find
\begin{eqnarray}
\label{PHA}
\Pi^{HA} &=& \frac{1}{\kappa} \Im m (\lambda_5\lambda^{*2}_6)\,
(\cos\theta\sin\beta\, -\, \sin\theta\cos\beta )\, ,\\
\label{PhA}
\Pi^{hA} &=& -\, \frac{1}{\kappa}\, \Im m (\lambda_5\lambda^{*2}_6)\,
(\sin\theta\sin\beta\, +\,\cos\theta\cos\beta )\, ,
\end{eqnarray}
where $\kappa$ is some squared mass combination which depends entirely
on the VEV's $v_1$, $v_2$, and $\Re e\lambda_5$ and $\Re e\lambda_6$.
In our analysis, we shall treat the CP-violating squared mass terms,
$\Pi^{HA}$ and $\Pi^{hA}$, as small parameters compared to the Higgs
particle masses, {\em i.e.} $\Pi^{HA}/M^2_H\ll 1$ and
$\Pi^{hA}/M^2_h\ll 1$. In Section \ref{sec:CPHA}, we will find that
this mass pattern is also compatible with experimental upper bounds on
EDM's.

\noindent (ii) {\em $HA$ mixing in a two-Higgs doublet model with
heavy Majorana neutrinos}. In this two-Higgs doublet model, the
discrete symmetry D is broken explicitly by the Majorana terms of two
isosinglet neutrinos, call them ${S_1}_R$ and ${S_2}_R$. In addition,
we introduce a sequential weak isodoublet in the model, $(\nu_4,
E)_L$. To avoid possible phenomenological limits coming from the
presently observed sector, we assume the complete absence of
inter-family mixings with the three light generations. The Yukawa
sector containing the neutrino mass matrix $M^\nu$ of our CP-violating
scenario reads:
\begin{equation}
\label{massnu}
-{\cal L}^\nu_Y\ =\ \frac{1}{2} \Big(\bar{\nu}_{4L},\
(\bar{S}_{1R})^C,\ (\bar{S}_{2R})^C \Big) \left[\begin{array}{ccc} 0 &
a & b \\ a & A & 0 \\ b & 0 & B \end{array} \right]\, \left(
\begin{array}{c} ({\nu_4}_L)^C\\ {S_1}_R\\ {S_2}_R
\end{array} \right)\ +\ \mbox{H.c.}
\end{equation}
In Eq.\ (\ref{massnu}), the parameters $A$ and $B$ can always be
chosen to be real, whereas $a$ and $b$ are in general complex. CP is
violated only if the rephasing invariant quantity $\Im m (ab^*) \not=
0$, which is only possible with the presence of two heavy singlets in
the model.  The symmetric mass matrix in Eq.\ (\ref{massnu}) can be
diagonalized by the unitary transformation: $U^T M^\nu U\, =\,
\widehat{M}^\nu$, where $\widehat{M}^\nu$ is a diagonal matrix
containing the physical heavy neutrino masses, $m_i$ ($i=1,2,3$). From
the three heavy Majorana neutrinos, denoted by $N_1$, $N_2$, and
$N_3$, that the model predicts, $N_1$ is predominantly a SU(2)$_L$
isodoublet, and $N_2$ and $N_3$ are mainly singlets in the limit of
$A,B\gg a,b$.  Moreover, the Lagrangians governing the interactions
between $N_i$ and $h$, $H$, $A$ are given by \cite{zpc}:
\begin{eqnarray}
\label{ANN}
{\cal L}_{A} &=& \frac{i g}{4M_W}\ A\, \chi^u_A\, \sum_{i,j=1}^{3}
\bar{N}_i \Big[ \gamma_5\, (m_i+m_j)\Re e C_{ij} +\ i(m_j-m_i)\Im m
C_{ij} \Big] N_j\, ,\\
\label{HNN}
{\cal L}_{h,H} &=& -\ \frac{g}{4M_W}\ (h\, \chi^u_h +\, H\, \chi^u_H )
                        \nonumber\\ &&\times\, \sum_{i,j=1}^{3}
                        \bar{N}_i \Big[ (m_i+m_j)\Re e C_{ij} +\
                        i\gamma_5 (m_j-m_i)\Im m C_{ij} \Big] N_j\, ,
\end{eqnarray} 
where $\chi^u_A=\cot\beta$, $\chi^u_h$ and $\chi^u_H$ are given in
Eq.\ (\ref{chis}), and $C_{ij}$ is a $3\times 3$ mixing matrix defined
as $C_{ij}\equiv U_{1i}U^*_{1j}$. Thus, the rephasing invariant and
CP-violating quantity mentioned above may equivalently be expressed as
\begin{equation} 
\label{CPneutr}
       \Im m C^2_{12}\ =\ \sin\delta_{CP} |C_{12}|^2\, .
\end{equation}
We shall assume that $\Im m C^2_{12}$ in Eq.\ (\ref{CPneutr}) takes
the maximum possible value, {\em i.e.}, it is of order one. Further
details on that model may be found in \cite{IKP}. Note that a
CP-violating $HA$ mixing may also be induced within the MSSM, in which
neutralinos and charginos may assume the r\^ole of heavy Majorana
neutrinos. In the MSSM, the SU(2)$_L\times$U(1)$_Y$-singlet bilinear
term $\mu$ and the tri-linear soft-SUSY-breaking couplings $A$ may
contain non-trivial CP-violating phases, which lead to complex
chargino- and neutralino-mass matrices \cite{KO}. Consequently,
interaction Lagrangians of the form given in Eqs.\ (\ref{ANN}) and
(\ref{HNN}) may naturally occur in the MSSM.

\begin{center}
\begin{picture}(200,80)(0,0)
\SetWidth{0.8}

\DashArrowLine(46,40)(84,40){5}\Text(60,49)[]{$H$}
\DashArrowLine(117,40)(154,40){5}\Text(140,49)[]{$A$}
\ArrowArc(100,40)(17,0,180)\Text(100,67)[]{$n_i$}
\ArrowArc(100,40)(17,180,360)\Text(100,13)[]{$n_j$}

\end{picture}\\[0.7cm]
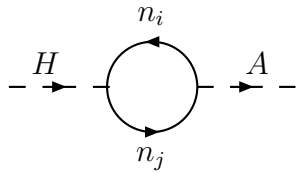
{\small {\bf Fig.\ 4:} $HA$ mixing induced by heavy Majorana
neutrinos.}
\end{center}

As shown in Fig.\ 4, the $HA$ mixing may be induced radiatively by
heavy Majorana neutrino loops. Taking the Lagrangians (\ref{ANN}) and
(\ref{HNN}) into account, we calculate the CP-violating $HA$ mixing in
our two-Higgs doublet model, {\em viz.}
\begin{equation}
\label{HZmix}
\frac{\widehat{\Pi}^{AH}(s)}{s}\ =\ -\frac{\alpha_w}{4\pi}\,
\chi^u_A\chi^u_H\, \sum\limits_{j>i}^{3} \Im m C^2_{ij}\,
\sqrt{\lambda_i\lambda_j}\, \Big[ B_0(s/M^2_W,\lambda_i,\lambda_j)\,
+\, 2B_1(s/M^2_W,\lambda_i,\lambda_j)\Big]\, ,
\end{equation}
where $\widehat{\Pi}^{AH}(s)=\widehat{\Pi}^{HA}(s)$, $\lambda_i =
m^2_i/M^2_W$, and $B_0$ and $B_1$ are the usual Veltman-Passarino loop
functions, evaluated in the conventions of Ref.\ \cite{BAK}. The
transition $G^0H$ is easily recovered from Eq.\ (\ref{HZmix}) by
setting $\chi^u_A=1$.  Correspondingly, the other CP-violating
transitions, $G^0h$ and $Ah$, may be obtained by replacing $\chi^u_H$
with $\chi^u_h$ in Eq.\ (\ref{HZmix}). All these CP-violating
self-energies are UV finite, and hence, do not require renormalization
(see also Appendix A on that matter).

Since there is an established equivalence of the one-loop PT $n$-point
correlation functions with those obtained in the background field
method for the gauge-fixing-parameter value $\xi_Q=1$ \cite{BFM,GPT},
we shall calculate the one-loop PT self-energies, using the latter
method. Furthermore, UV divergences occuring in the dispersive parts
of the self-energies are absorbed by mass and wave-function
renormalization constants. However, near the resonant region,
$s\approx M^2_H\approx M^2_A$, OS renormalization renders the
dispersive self-energies, $\Re e\widehat{\Pi}^{HH}(s)$ and $\Re
e\widehat{\Pi}^{AA}(s)$, very small of order $g^6$ ({\em cf.}\ Eq.\ 
(\ref{ReHH})), so that these high-order terms can be neglected.
Therefore, only one-loop absorptive $HH$ and $AA$ self-energies are of
interest here. Such an approach may be viewed as an improved Born
approximation. It is then straightforward to obtain for the different
decay modes
\begin{eqnarray}
\label{Piff}
\Im m\widehat{\Pi}^{HH}_{(f\bar{f})}(s) &=& \frac{\alpha_w N^f_c}{8}
(\chi^f_H)^2\, s\, \frac{m^2_f}{M^2_W}
\left(1-\frac{4m^2_f}{s}\right)^{3/2} \theta (s-4m^2_f )\, ,\\
\label{PiAff}
\Im m\widehat{\Pi}^{AA}_{(f\bar{f})}(s) &=& \frac{\alpha_w N^f_c}{8}
(\chi^f_A)^2\, s\, \frac{m^2_f}{M^2_W}
\left(1-\frac{4m^2_f}{s}\right)^{1/2} \theta (s-4m^2_f )\, ,\\
\label{PiVV}
\Im m\widehat{\Pi}^{HH}_{(VV)}(s) &=&
\frac{n_V\alpha_w}{32}(\chi^V_H)^2 \frac{M^4_H}{M^2_W}
\left(1-\frac{4M^2_V}{s}\right)^{1/2}\nonumber\\ &&\times \left[
1+4\frac{M^2_V}{M^2_H}-4\frac{M^2_V}{M^4_H}(2s-3M^2_V) \right] \theta
(s-4M^2_V )\, .
\end{eqnarray}
Here, $\alpha_w=g^2/4\pi$, $n_V=2$, 1 for $V\equiv W$, $Z$,
respectively, and $N^f_c=1$ for leptons and 3 for quarks. In Eqs.\
(\ref{Piff})--(\ref{PiVV}), $\chi^f_{H,A}$ are the model-dependent
factors in Eq.\ (\ref{chis}), and $\chi^{W}_{h,H}=\chi^{Z}_{h,H}$ are
similar factors that multiply the couplings $HWW$ and $HZZ$ of the
SM. The explicit dependence of $\chi^{W}_{h,H}$ on the angles $\theta$
and $\beta$ is given by
\begin{equation}
\label{chiWZ}
\chi^{W}_h\, =\, \cos\theta\cos\beta\, +\, \sin\theta\sin\beta\,
,\quad \chi^{W}_H\, =\, -\sin\theta\cos\beta\, +\,
\cos\theta\sin\beta\, .
\end{equation}
In addition, one has $\chi^V_A=0$ because of CP invariance. Thus, only
fermions can contribute to $\Im m \widehat{\Pi}^{AA}(s)$. Other
channels involving the $HZA$ vertex may also give contributions to the
absorptive self-energies.  However, for the kinematic range of our
interest, $M_H\simeq M_A$, relevant for resonant CP violation, these
absorptive channels are considered to be phase-space suppressed and
hence have not been taken into account.

\setcounter{equation}{0}
\section{\label{sec:CPHA} Resonant CP violation through {\boldmath
$HA$}  mixing}

In this section, we shall study CP-violating phenomena which are
induced by resonant transitions of a CP-even Higgs particle, $H$, into
a CP-odd Higgs scalar, $A$. The importance of the resonant enhancement
of CP violation through $HA$ particle mixing has been addressed in an
earlier communication \cite{APRL}. Here, we shall present further
details of the calculation and analyze possible low-energy constraints
on $HA$ mixing, such as bounds coming from electric dipole moments
(EDM's) of neutron, electron and muon.

The most ideal place to look for resonant CP-violating $HA$
transitions is at $e^+e^-$ and, most interestingly, at muon colliders
\cite{mumu}.  In general, there are many observables suggested at
high-energy colliders \cite{DV,NP1,Gavela,Nacht,Peskin,WB} that may be
formed to project out different CP/T-noninvariant contributions. All
the CP-violating observables, however, may fall into two categories,
depending on whether they are even or odd under naive CPT
transformations. For instance, typical CP-odd and CPT-even observables
in a process, such as $p\bar{p}\to t\bar{t} X$ \cite{DV,NP1} or
$e^+e^-\to t\bar{t} X$ \cite{PN}, are triple-product correlations of
the type, {\em e.g.}, $\langle \vec{k}_p \cdot \vec{k}_t \times
\vec{k}_{\bar{t}} \rangle$ or $\langle \vec{k}_e \cdot \vec{k}_t
\times \vec{k}_{\bar{t}} \rangle $, based on the three-momenta of the
initial beam particles $p$ or $e$ and top quarks in the final
state. In this class of observables, CP violation may occur already
in the Born approximation \cite{DV,NP1,Gavela}.  The other category
comprises CP- and CPT-odd observables of the form, $\langle \vec{s}_t
\vec{k}_t \rangle$ or $\langle \vec{s}_{\bar{t}}
\vec{k}_{\bar{t}}\rangle $ \cite{PN,Peskin,CK}, where $\vec{s}_t$ and
$\vec{k}_t$ are respectively the spin and the three-momentum of the
top-quark in the $t\bar{t}$ centre of mass (c.m.) system.  These
observables, being odd under naive CPT transformations, can only
combine with absorptive loops, which are also naively odd under CPT,
to produce a real and CPT-invariant contribution to the matrix element
squared. 

Our quantitative analysis of $HA$ mixing phenomena will rely on
CP-violating quantities of the second class, mentioned above.  For
definiteness, assuming that having longitudinally polarized muon beams
will be feasible without much loss of luminosity, we shall consider
the CP asymmetry \cite{CPmumu}
\begin{equation}
\label{CPmuobs}
{\cal A}^{(\mu)}_{CP}\ =\ \frac{\sigma (\mu^-_L\mu^+_L\to f\bar{f})\
-\ \sigma (\mu^-_R\mu^+_R\to f\bar{f})}{\sigma (\mu^-_L\mu^+_L\to
f\bar{f})\ +\ \sigma (\mu^-_R\mu^+_R\to f\bar{f})}\ .
\end{equation}
If one is able to tag on the final fermion pair $f\bar{f}$ ({\em
e.g.}, $\tau^+\tau^-$, $b\bar{b}$, or $t\bar{t}$), ${\cal A}^{(\mu
)}_{CP}$ is then a genuine observable of CP violation, as the helicity
states $\mu^-_L\mu^+_L$ transform into $\mu^-_R\mu^+_R$ under CP in
the c.m.\ system. Similarly, at $e^+e^-$ or $p\bar{p}$ machines, one
can define the CP asymmetry
\begin{equation}
\label{CPelobs}
{\cal A}^{(e)}_{CP}\ =\ \frac{\sigma (e^-e^+\to f_L\bar{f}_L\, X)\ -\
\sigma (e^-e^+\to f_R\bar{f}_R\, X)}{\sigma (e^-e^+\to f_L\bar{f}_L\,
X)\ +\ \sigma (e^-e^+\to f_R\bar{f}_R\, X)}\ ,
\end{equation}
and an analogous observable ${\cal A}^{(p)}_{CP}$ involving $p\bar{p}$
beams. In Eq.\ (\ref{CPelobs}), the chirality of fermions, such as the
top quark, may not be directly observed. However, the decay
characteristics of a left-handed top quark differ substantially from
those of its right-handed component, giving rise to distinct
angular-momentum distributions and energy asymmetries of the produced
charged leptons and jets \cite{Peskin,WB}.

Since our main interest is in resonant $HA$ or $hA$ transitions, we
shall assume for simplicity that only one CP-even Higgs particle, $H$
say, has a mass quite close to $A$, {\em i.e.}, $M_H - M_A \ll M_H,
M_A$, while the other CP-even Higgs, $h$, is much lighter than $H$,
$A$, and vice versa.  As has been noted in Section \ref{sec:RCPV}, $h$
will effectively decouple from the mixing system, having negligible
contributions to both cross section and CP asymmetry at c.m.\ energies
$s\approx M_H,M_A$. This is also the main reason accounting for the
fact that CP violation through $HZ$ ($G^0H$) mixing has been found to
be small \cite{APRL}, as $H$ only couples to the longitudinal
component of the $Z$ boson, the massless would-be Goldstone $G^0$. To
give an estimate, in the two-Higgs model with heavy Majorana
neutrinos, we find that ${\cal A}^{(\mu )}_{CP}\approx 2.\ 10^{-2}$
for $M_H=500$ GeV and $m_{1,2,3} = 0.5,\ 1.5,\ 3$ TeV, while the
production cross-section is $\sigma \simeq 1$ fb. It is therefore
unlikely to observe $HZ$-mixing effects, even if one assumes a high
integrated luminosity of 50 fb$^{-1}$, designed for $e^+e^-$ and
$\mu^+\mu^-$ colliders.

We shall now focus our attention on the resonant transition amplitudes
${\cal T}_L (\mu^+_L\mu^-_L \to H^*,\, A^* \to f\bar{f})$ and ${\cal
  T}_R (\mu^+_R\mu^-_R \to H^*,\, A^* \to f\bar{f})$.  Therefore, it
will prove useful to further decompose these amplitudes as follows:
\begin{eqnarray}
\label{TLRHA}
{\cal T}_L &=& {\cal T}^H_L (\mu^+_L\mu^-_L \to H^*H^*, A^*H^* \to f\bar{f})\
+\ {\cal T}^A_L (\mu^+_L\mu^-_L \to A^*A^*,H^*A^* \to f\bar{f})\, ,\nonumber\\
{\cal T}_R &=& {\cal T}^H_R (\mu^+_R\mu^-_R \to H^*H^*, A^*H^* \to f\bar{f})\
+\ {\cal T}^A_R (\mu^+_R\mu^-_R \to A^*A^*,H^*A^* \to f\bar{f})\, .
\end{eqnarray}
In Eq.\ (\ref{TLRHA}), the individual amplitudes ${\cal T}^H_{L,R}$
(${\cal T}^A_{L,R}$) are uniquely specified by the tree-level coupling
of the virtual Higgs particle $H$ ($A$) to the fermions $f$ in the
final state. In this way, we can evaluate the following CP-violating
differences of the matrix elements squared:
\begin{eqnarray}
\label{TCPA}
\Delta {\cal T}^A &=& |{\cal T}^A_L|^2-|{\cal T}^A_R|^2 \nonumber\\
&=& -4\frac{r_d\, \Im m\widehat{\Pi}^{HH}(s)}{\widehat{\Pi}^{AH}(s)}\,
|\hat{\Delta}_{AH}(s)|^2\, |{\cal M}(A^*\to \mu^-_L\mu^+_L)|^2\,
|{\cal M}(A^*\to f\bar{f})|^2\, ,\\
\label{TCPH}
\Delta {\cal T}^H &=& |{\cal T}^H_L|^2-|{\cal T}^H_R|^2 \nonumber\\
&=& 4\frac{\Im m\widehat{\Pi}^{AA}(s)}{r_d\, \widehat{\Pi}^{AH}(s)}\,
|\hat{\Delta}_{AH}(s)|^2\, |{\cal M}(H^*\to \mu^-_L\mu^+_L)|^2\,
|{\cal M}(H^*\to f\bar{f})|^2\, ,
\end{eqnarray}
where $r_d=\chi^d_H/\chi^d_A$ and $\widehat{\Pi}^{AH}(s)$ is assumed
to be real (dispersive), for the two-Higgs doublet models discussed in
Section \ref{sec:HA}.  In Eqs.\ (\ref{TCPA}) and (\ref{TCPH}), the tree-level
squared amplitudes involving the $Hf\bar{f}$ and $Af\bar{f}$ couplings
obey the relation: $|{\cal M}(H^* \to f\bar{f})|^2 = x_f(s)\, |{\cal
  M}(A^* \to f\bar{f})|^2$, where $x_f(s) = r^2_f (1-4m^2_f/s)$ with
$r_f=\chi^f_H/\chi^f_A$.  Furthermore, the CP-conserving squared
amplitudes are given by
\begin{eqnarray}
\label{TA2}
|{\cal T}^A|^2 &=& |{\cal T}^A_L|^2+|{\cal T}^A_R|^2 \ =\ 2\left[\,
r^2_d\, +\, \frac{(s-M^2_H)^2+(\Im
m\widehat{\Pi}^{HH}(s))^2}{(\widehat{\Pi}^{AH}(s))^2} \, \right]
\nonumber\\ &&\times\, |\hat{\Delta}_{AH}(s)|^2\, |{\cal M}(A^*\to
\mu^-_L\mu^+_L)|^2\, |{\cal M}(A^*\to f\bar{f})|^2\, ,\\
\label{TH2}
|{\cal T}^H|^2 &=& |{\cal T}^H_L|^2+|{\cal T}^H_R|^2 \ =\ 2\left[\,
r^{-2}_d\, +\, \frac{(s-M^2_A)^2+(\Im
m\widehat{\Pi}^{AA}(s))^2}{(\widehat{\Pi}^{AH}(s))^2} \,
\right]\nonumber\\ &&\times\, |\hat{\Delta}_{AH}(s)|^2\, |{\cal
M}(H^*\to \mu^-_L\mu^+_L)|^2\, |{\cal M}(H^*\to f\bar{f})|^2\, .
\end{eqnarray}
Considering Eqs.\ (\ref{TCPA})--(\ref{TH2}), it is not difficult to
calculate the CP asymmetry
\begin{eqnarray}
\label{ACPmu}
{\cal A}^{(\mu )}_{CP}(s) &=&\frac{\Delta {\cal T}^A\, +\, \Delta
{\cal T}^H}{ |{\cal T}^A|^2\, +\, |{\cal T}^H|^2} \\ && \hspace{-45pt}
=\frac{ 2\, r_d\, \widehat{\Pi}^{AH}\, (x_f \Im m\widehat{\Pi}^{AA} \,
- \, \Im m\widehat{\Pi}^{HH} ) }{ x_f\, r^2_d [(s-M^2_A)^2 + (\Im
m\widehat{\Pi}^{AA})^2] + x_f(\widehat{\Pi}^{AH})^2 + (s-M^2_H)^2 +
(\Im m\widehat{\Pi}^{HH})^2 + (r_d\widehat{\Pi}^{AH})^2}\, .
\nonumber
\end{eqnarray}
The analytic result of ${\cal A}^{(\mu )}_{CP}(s)$ in Eq.\
(\ref{ACPmu}) simplifies to the qualitative estimate presented in
\cite{APRL}, if finite mass effects of the asymptotic states are
neglected, the value $r_f\approx 1$ is taken, ({\em i.e.}, $x_f\approx
1$), and $(\widehat{\Pi}^{AH})^2$ terms are omitted in the
CP-conserving part of the squared amplitude. In the present analysis,
we shall include the high-order $(\widehat{\Pi}^{AH})^2$ terms and
take into account all those refinements inherent to the two-Higgs
doublet model.

In order to reduce the large number of independent parameters that may
vary independently, we fix the angles $\beta$ and $\theta$ to the
values such that $\tan\beta=2$ and $\tan\theta=1$. In this scheme, the
heaviest CP-even Higgs, $H$, has a significant coupling to fermions,
{\em i.e.}, $\chi^W_H=\chi^Z_H\ll \chi^u_H, \chi^u_A$, whereas
$(\chi^W_H)^2= (\chi^Z_H)^2 \approx 0.1$. For $M_A > 2 M_Z$, such a
scheme is motivated by the MSSM and leads to nearly degenerate $H$ and
$A$ scalars, {\em i.e.}, $M_H \approx M_A$ \cite{JFG}. This mass
relation also turns out to be very typical within SUSY unified models
\cite{SUSYGUT}.  Nevertheless, we shall not consider here that $M_H$
is very strongly correlated with $M_A$.  In the two-Higgs doublet
model with heavy Majorana neutrinos, the three heavy neutrino masses
are also fixed to $m_1=0.5$ TeV, $m_2=1$ TeV and $m_3=1.5$ TeV. In the
two-Higgs doublet model with broken D symmetry, the tree-level $HA$ or
$hA$ mixings, $\Pi^{HA}$ and $\Pi^{hA}$, are considered to be small
phenomenological parameters, compared to the squared masses of all
three Higgs particles, which is a result of constraints coming from
the neutron EDM, as we will briefly discuss at the end of this
section.  For the top quark mass, we use $m_t=170$ GeV close to its
experimental mean value \cite{PDG}.

Assuming that tuning the collider c.m.\ energy to the mass of $H$ or
$h$ is possible, {\em i.e.}, $\sqrt{s}=M_H$ or $M_h$, we analyze the
following two reactions:
\begin{eqnarray}
\mbox{(a)} && \mu^+_L\mu^-_L \to h^*,A^* \to b\bar{b}\, ,\quad
\mbox{with}\ M_A=170\ \mbox{GeV}\, ,\nonumber\\ \mbox{(b)} &&
\mu^+_L\mu^-_L \to H^*,A^*\to t\bar{t}\, ,\quad \mbox{with}\ M_A=400\
\mbox{GeV}\, . \nonumber
\end{eqnarray}
In Fig.\ 5(a), we display cross sections (solid lines) and CP
asymmetries (dotted lines) as a function of the c.m.\ energy,
$\sqrt{s}$, for the two reactions (a) and (b) in the two-Higgs doublet
model with heavy Majorana neutrinos.  In reaction (a), one can have a
significant CP-violating signal for $M_H\gg M_h$ and $M_A$, if
$M_h=170\pm 8$ GeV.  As can also be seen from Fig.\ 5(a), we observe
the resonant enhancement of ${\cal A}^{(\mu)}_{CP} (M^2_h)$ when
$M_h\approx M_A$.  CP violation in reaction (b) may be probed for a
wider range of Higgs-boson masses, {\em i.e.}, for $M_H=380$ GeV $-$
420 GeV.  According to the discussion in Section \ref{sec:RCPV}, CP
violation becomes maximal when the necessary condition (\ref{CPN1})
for $M_H\approx M_A$ is met. Furthermore, CPT symmetry prescribes the
constraint for the resonant cross sections ({\em cf.}\ Eq.\ 
(\ref{CPTcros}))
\begin{equation}
\label{CPTmu}
\sigma (\mu^+_L\mu^-_L \to H^*(h^*),A^*\to \mbox{all})\ =\ \sigma
(\mu^+_R\mu^-_R \to H^*(h^*),A^*\to \mbox{all})\, ,
\end{equation}
which may generally reduce the magnitude of CP violation and so
influence the actual dependence of $A_{CP}$ on the $HA$ or $hA$
mixing. However, our resummation approach takes automatically account
of such CPT constraints, as has been explicitly demonstrated in
Section \ref{sec:CPT} ({\em cf.}\ Eq.\ (\ref{aCP1})).  In Fig.\ 5(b),
we show how $|A_{CP}|$ varies as a function of the parameter
$x_A=\Pi^{HA}/\Im m(\widehat{\Pi}^{HH}-\widehat{\Pi}^{AA})$ or
$\Pi^{hA}/\Im m(\widehat{\Pi}^{hh}-\widehat{\Pi}^{AA})$, in the
two-Higgs doublet model with a D-symmetry breaking. We consider the
kinematic region for resonantly enhanced CP violation, {\em i.e.},
$M_A=M_h$ ($M_H$) for the reaction (a) (reaction (b)). We find that
CP-violating effects could become very large, if the parameter $x_A$
was tuned to the value $x_A=1$ for the process (a) and $x_A=3$ for the
process (b).
  
At the next linear $e^+e^-$ colliders (NLC's), Higgs bosons may
copiously be produced either via the Bjorken process for c.m.\ energies
up to 0.5 TeV or through $WW$ fusion at higher energies \cite{NLC}.
The most convenient way is to study CP violation in the kinematic
range of the Higgs production and decay \cite{PN,BG}. Therefore, we
shall be interested in the observable formed by differential cross
sections
\begin{equation}
\label{CPeldif}
\overline{{\cal A}}^{(e)}_{CP}(\hat{s})\ =\ \frac{d\sigma (e^-e^+\to
f_L\bar{f}_L\, X)/d\hat{s}\ -\ d\sigma (e^-e^+\to f_R\bar{f}_R\,
X)/d\hat{s}}{ d\sigma (e^-e^+\to f_L\bar{f}_L\, X)/d\hat{s}\ +\
d\sigma (e^-e^+\to f_R\bar{f}_R\, X)/d\hat{s}}\ ,
\end{equation}
where $\hat{s}$ is the invariant mass energy of the produced final
fermions $f$. As final states, we may take bottom or top quarks.
Since $A$ does not couple to $WW$ or $ZZ$, the CP asymmetry
$\hat{{\cal A}}^{(e)}_{CP}$ in Eq.\ (\ref{CPeldif}) takes the simple
form
\begin{equation}
\label{ACPel}
\overline{{\cal A}}^{(e)}_{CP}(\hat{s}) \ =\ -\, \frac{
2r_f\widehat{\Pi}^{AH}(\hat{s})\, \Im m\widehat{\Pi}^{AA}(\hat{s})}{
r^2_f\, [(\hat{s}-M^2_A)^2 + (\Im m\widehat{\Pi}^{AA}(\hat{s}))^2] +
(\widehat{\Pi}^{AH}(\hat{s}))^2}\, .
\end{equation}
In Fig.\ 6(a), we display the dependence of $\overline{{\cal
A}}^{(e)}_{CP}$ as a function of $\hat{s}$. As expected, we find a
resonant enhancement of CP violation when $M_A\approx M_h$ or
$M_H$. Since the destructive term, $\Im m\widehat{\Pi}^{HH}$, is
absent in Eq.\ (\ref{ACPel}), CP violation may become even larger,
{\em i.e.}, of order unity for specific values of the parameter
$x_A$. Indeed, we see from Fig.\ 6(b) that $\overline{{\cal
A}}^{(e)}\approx 1$, if $x_A=0.07$ (3) for the reaction with
longitudinally polarized $b$ ($t$) quarks in the final state.
Estimates of the CP asymmetry based on totally integrated cross
sections, ${\cal A}^{(e)}_{CP}$ in Eq.\ (\ref{CPelobs}), are presented
in \cite{PN}.  The authors \cite{PN} find that ${\cal A}^{(e)}_{CP} <
15\%$ for Higgs masses $M_H<600$ GeV and cross sections $\sigma
(e^-e^+\to H^* (h^*), A^*\to t\bar{t}\, X)\approx 10-100$ fb for 
c.m.\ energies of 1 -- 2 TeV. Such CP-violating effects have high chances 
to be detected at future NLC's.
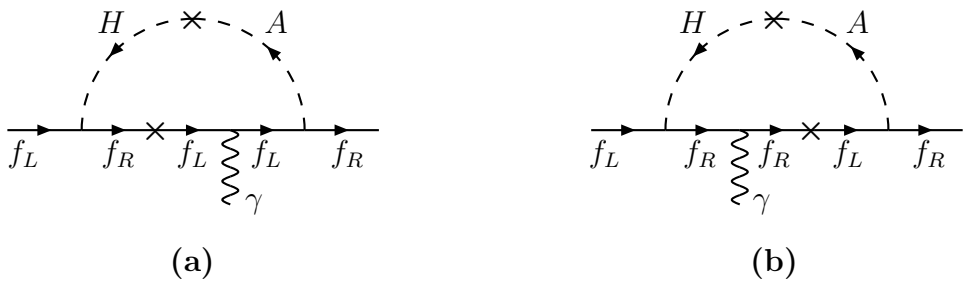
\begin{center}
\begin{picture}(360,120)(0,0)
\SetWidth{0.8}

\ArrowLine(0,50)(28,50)\Text(0,41)[l]{$f_L$}
\ArrowLine(28,50)(56,50)\Text(42,41)[]{$f_R$}
\ArrowLine(56,50)(84,50)\Text(70,41)[]{$f_L$}
\ArrowLine(84,50)(112,50)\Text(98,41)[]{$f_L$}
\ArrowLine(112,50)(140,50)\Text(135,41)[r]{$f_R$}
\DashArrowArc(70,50)(42,0,90){5}\Text(34,87)[lb]{$H$}
\DashArrowArc(70,50)(42,90,180){5}\Text(106,87)[rb]{$A$}
\Photon(84,22)(84,50){3}{4}\Text(90,22)[l]{$\gamma$}

\Text(56,50)[]{{\boldmath $\times$}} \Text(70,92)[]{{\boldmath
$\times$}} \Text(70,0)[]{{\bf (a)}}


\ArrowLine(220,50)(248,50)\Text(220,41)[l]{$f_L$}
\ArrowLine(248,50)(276,50)\Text(262,41)[]{$f_R$}
\ArrowLine(276,50)(304,50)\Text(290,41)[]{$f_R$}
\ArrowLine(304,50)(332,50)\Text(318,41)[]{$f_L$}
\ArrowLine(332,50)(360,50)\Text(355,41)[r]{$f_R$}
\DashArrowArc(290,50)(42,0,90){5}\Text(254,87)[lb]{$H$}
\DashArrowArc(290,50)(42,90,180){5}\Text(326,87)[rb]{$A$}
\Photon(276,22)(276,50){3}{4}\Text(282,22)[l]{$\gamma$}

\Text(304,50)[]{{\boldmath $\times$}} \Text(290,92)[]{{\boldmath
$\times$}} \Text(290,0)[]{{\bf (b)}}

\end{picture}\\[0.7cm]
{\small {\bf Fig.\ 7:} Diagrams contributing to the EDM of a fermion.}
\end{center}

At the LHC, the respective CP asymmetry $\overline{{\cal
    A}}^{(p)}_{CP} (\hat{s})$ may be obtained from Eq.\ (\ref{ACPel}),
for Higgs particles that have a production mechanism similar to that
at the NLC. If the Higgs is produced via gluon fusion, one has to use
the analytic expression of ${\cal A}^{(\mu )}_{CP}(\hat{s})$ and
replace $r_d$ with $r_u$ in Eq.\ (\ref{ACPmu}). Unless CP violation is
resonantly amplified, {\em i.e.}, of order one, the chances to detect
CP-violating phenomena on the Higgs-resonance line after removing the
contributing background appear to be quite limited at the LHC. It is
therefore worth stressing that a large CP-violating signal at the
Higgs-boson peak will certainly point towards the existence of an
almost degenerate $HA$ mixing system. 
\begin{center}
\begin{picture}(360,160)(0,0)
\SetWidth{0.8}

\ArrowLine(110,30)(145,60)\Text(130,35)[lb]{$f_L$}
\DashLine(145,60)(180,100){5}\Text(146,68)[rb]{$A$}
\Text(162,86)[rb]{$H$}\Text(160,77)[]{{\boldmath $\times$}}
\Photon(180,100)(215,60){3}{5}\Text(200,85)[l]{$\gamma,Z$}
\ArrowLine(145,60)(215,60)\Text(180,51)[l]{$f_R$}
\ArrowLine(215,60)(250,30)\Text(218,35)[lb]{$f_R$}
\Photon(180,100)(180,150){3}{5}\Text(186,140)[l]{$\gamma$}
\GCirc(180,100){10}{0.5}

\end{picture}\\[0.7cm]
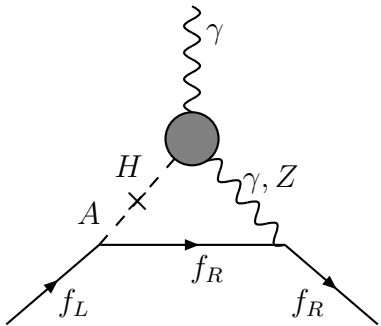
{\small {\bf Fig.\ 8:} The two-loop Barr-Zee mechanism for generating
EDM.}
\end{center}

The presence of a $HA$ operator may also contribute to other
low-energy CP-violating observables. CP-violating quantities sensitive
to $HA$ terms, for which there exist quite strict experimental upper
bounds, are the EDM's of the neutron, electron and muon.  From the
one-loop flavour diagrams shown in Fig.\ 7, it is straightforward to
calculate
\begin{equation}
\label{EDMf}
d_f/e\ \approx\ -Q_f \frac{\alpha_w}{4\pi}\, \frac{m_f}{M^2}\,
\frac{m^2_f}{M^2_W}\, \xi_{HA}\, \Big[\,
\ln\Big(\frac{m^2_f}{M^2}\Big)\, +\, \frac{3}{2}\, \Big]\, ,
\end{equation}
with $M=(M_H +M_A)/2$, $Q_f$ denoting the charge of fermion in units
of $|e|$ and
\begin{equation}
\label{xiHA}
\xi_{HA}\ =\ \chi^f_A\chi^f_H\, \frac{\widehat{\Pi}^{HA}(M^2)}{M^2}\ .
\end{equation}
The most severe constraint comes from the EDM of the neutron, which
has the experimental upper bound $(d_n/e) < 1.1\ 10^{-25}$ cm, at
95$\%$ of confidence level (CL) \cite{PDG}. The contribution of the
one-loop graphs in Fig.\ 7 is much smaller. Taking the typical values
of $m_d=10$ MeV and $M_W\approx M\approx 100$ GeV, Eq.\ (\ref{EDMf})
predicts $(d_n/e) < 3.\ 10^{-30}\, \xi_{HA}$ cm, far beyond the
experimental bound. Nevertheless, the two-loop Barr-Zee (BZ) mechanism
\cite{BZ} shown in Fig.\ 8 may have a significant impact on the actual
size of the EDM, thus leading to tighter bounds on the $HA$ mixing
parameter $\xi_{HA}$. In general, the theoretical prediction for the EDM
is enhanced by the absence of chirality suppressed terms of order
$m^2_f/M^2_W$, despite the fact that the BZ mechanism occurs at the
two-loop order. Thus, the net effect is to increase the value of
$(d_n/e)$ in Eq.\ (\ref{EDMf}) by a factor $\alpha (M^2_W/m^2_d)
\approx 10^6$, where $\alpha = 1/137$ is the fine-structure constant
\cite{BZ}. A recent analysis of constraints from the electron and
neutron EDM's is presented in \cite{EDMnew} for the two-Higgs doublet
model with maximal CP violation, where the Weinberg's unitarity bound
is almost saturated \cite{SW}.  Even though the prediction for the
electron EDM is below the experimental limit, the neutron EDM gives
theoretical bounds that may be evaded if the mass difference, $\Delta
M$, \ between $A$ and one of the CP-even Higgs scalars $H$ (or $h$) is
sufficiently small. In this way, the authors of \cite{EDMnew} find
\begin{equation}
\label{EDMHA}
\frac{\Delta M}{M}\ \approx\ \frac{\widehat{\Pi}^{HA}}{M^2}\ <\ 0.10,\
0.13,\ 0.24\, ,
\end{equation}
for $M=200,\ 400,\ 600$ GeV, respectively. Note that the constraints
on $\Delta M/M$, obtained in $\cite{EDMnew}$ for the diagonal Higgs-mass
basis, have been adapted to our off-diagonal $HA$ basis, to leading
order in $(M^2_H-M^2_A)/M^2$. The limits due to the neutron EDM
in Eq.\ (\ref{EDMHA}) are automatically satisfied by our new-physics
scenarios.  Of course, for large $\tan\beta$ values, {\em e.g.},
$\tan\beta > 3$, the above limits on the mass difference between $H$
and $A$ will be much weaker.  As has been discussed above however, our
resonant CP-violating phenomena through particle mixing can still be
very large as soon as the necessary conditions (\ref{CPN1}) and/or
(\ref{CPN2}) are fulfilled.

\setcounter{equation}{0}
\section{\label{sec:Fer} Fermionic case}

We shall consider CP violation induced by the mixing of two fermions.
Issues of fermionic mixing renormalization are discussed in Ref.\
\cite{BK&AP} and hence will not be repeated here. For our
illustrations, we shall assume the mixing of two Dirac particles,
which have a small mass difference compared to their masses. To be
specific, one may think of scatterings involving the mixing of a top
quark with a new sequential up-type fermion, $t'$, as shown in Fig.\
9. Such resonant transitions were also discussed in \cite{AP1}. Here, 
we follow a theoretically more rigorous approach.

Assuming that box or non-resonant graphs are very small, we can write
down the matrix element for our prototype transition, $d \phi^+\to
t^*, t'^*\to s \phi^+$, in the following compact form:
\begin{equation}
\label{Tds}
{\cal T}\ \approx\ {\cal T}^{res}_{ds}\ =\ V^s_i\, \left(\frac{1}{
\not\! p\, -\, {\cal H}(\not\! p ) }\right)_{ij} V^d_j\, ,
\end{equation}
where $V^{d}_i$ and $V^{s}_i$ are production and decay vertices
involving the quarks $t,\ t'$ and the charged scalar $\phi^+$.  The
propagator matrix $[\not\! p\, -\, {\cal H}(\not\! p )]^{-1}$,
which is defined in the sub-space formed by $t$ and $t'$, describes
the resonant dynamics of the $tt'$-mixing system. Its explicit form
will be given in this section later on. The scalar field $\phi^+$ may
represent either a physical Higgs boson in new-physics scenarios or
the would-be Goldstone boson of $W^+$, $G^+$, which is a good
approximation for the longitudinal $W$ boson at high energies.  Also,
the CP-conjugate process is given by
\begin{equation}
\label{TCPds}
{\cal T}^{CP}\ \approx\ \overline{{\cal T}}^{res}_{ds}\ =\
\overline{V}^s_i\, \left(\frac{1}{ \not\! p\, -\, \overline{{\cal
H}}(\not\! p ) }\right)_{ij} \overline{V}^d_j\, ,
\end{equation}
where $\overline{V}^{d,s}_i$ and $\overline{{\cal H}}(\not\!\! p)$ are
the CP transforms of ${V}^{d,s}_i$ and ${\cal H}(\not\!\! p)$,
respectively.  These CP transformations are very analogous to those
given in Eqs.\ (\ref{Vamp}) and (\ref{CPtrH}) for the scalar case. 
\begin{center}
\begin{picture}(400,110)(0,0)
\SetWidth{0.8}

\ArrowLine(20,20)(42,42)\Text(8,12)[lb]{$d_i$}
\DashArrowLine(20,80)(42,58){5}\Text(15,78)[rb]{$\phi^+$}
\GCirc(50,50){10}{0.5}
\ArrowLine(60,50)(90,50)\Text(75,61)[]{$t,t^{'}$}
\GCirc(100,50){10}{0.5}
\ArrowLine(110,50)(140,50)\Text(125,61)[]{$t,t^{'}$}
\DashArrowLine(158,58)(180,80){5}\Text(188,78)[lb]{$\phi^{+}$}
\ArrowLine(158,42)(180,20)\Text(188,12)[lb]{$d_j$}
\GCirc(150,50){10}{0.5}

\Text(100,0)[b]{{\bf (a)}}


\DashArrowLine(240,64)(280,64){5}\Text(240,75)[]{$\phi^+$}
\ArrowLine(240,36)(280,36)\Text(240,20)[lb]{$d_i$}
\DashArrowLine(300,64)(340,64){5}\Text(340,75)[]{$\phi^+$}
\ArrowLine(300,36)(340,36)\Text(345,20)[rb]{$d_j$}
\GOval(290,50)(28,20)(0){0.5}

\Text(290,0)[b]{{\bf (b)}}

\end{picture}\\[0.7cm]
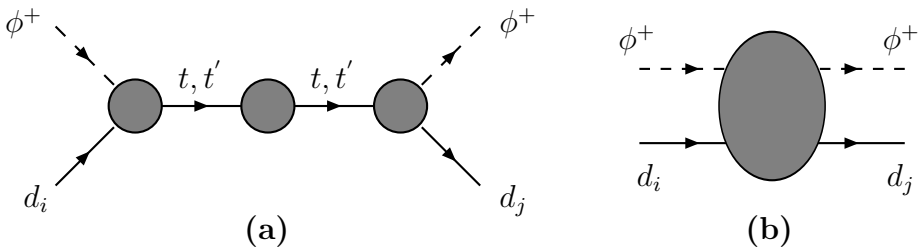
{\small {\bf Fig.\ 9:} Resonant CP-violating $tt'$ transitions.}
\end{center}

It is now worth emphasizing that we consider CP violation that
originates mainly from the mixing of the intermediate up-type
fermionic states $t$ and $t'$.  However, one should bear in mind that
the tree-level production and decay vertices $\overline{V}^{d,s}_i$,
{\em e.g.}, $Wu_id_j$, contain extra flavour rotations which result
from diagonalizing the mass matrix of the up-quark family by means of
bi-unitary transformations. In the SM, these rotations are
parameterized by the well-known Cabbibo-Kobayashi-Maskawa (CKM) mixing
matrix in the flavour basis where the down-quark mass matrix is
diagonal \cite{review}.  Therefore, one must always think of these CKM
rotations as being part of the fermionic effective Hamiltonian ${\cal
  H}(\not\! p)$. Besides the afore-mentioned CKM rotations, the
tree-level couplings $\phi^+ u_id_j$ depend on the up- and/or
down-family mass matrices, which can also introduce CP violation in
the vertices.  Formally speaking, such a CP violation is of
$\varepsilon'$ type and may therefore be competitive with our
$\varepsilon$-type effects.  Since it may be difficult to disentangle
the $\varepsilon$-type mass parameters from the $\varepsilon'$-type
couplings, we shall include both contributions in the Born
approximation and ignore possible high order $\varepsilon'$-type
effects in $\overline{V}^{d,s}_i$, as they are expected to be quite
small near the resonant region.

Within the context of our resummation approach, the propagator matrix
for two fermions, $\widehat{S}_{ij}(\not\! p)$, results from summing
up PT self-energies, $\widehat{\Sigma}_{ij}(\not\! p)$, {\em i.e.},
\begin{eqnarray}
\label{InvS}
[\not\! p - {\cal H}(\not\! p)]^{-1} &\equiv & \widehat{S}_{ij}(\not\!
p)\ =\ \delta_{ij} S_i^0 (\not\! p)\, -\, S_i^0(\not\! p)
\widehat{\Sigma}_{ij}(\not\! p) S_j^0 (\not\! p)\, +\,
\dots\nonumber\\ &=& \big[ S^{0-1}(\not\! p) + \widehat{\Sigma}(\not\!
p)\big]^{-1} \, =\, \left[ \begin{array}{cc} \not\! p - m_t +
\widehat{\Sigma}_{tt}(\not\! p) & \widehat{\Sigma}_{tt'}(\not\! p)\\
\widehat{\Sigma}_{t't}(\not\! p) & \not\! p - m_{t'} +
\widehat{\Sigma}_{t't'}(\not\! p)
\end{array} \right]^{-1},\qquad
\end{eqnarray}
where
\begin{eqnarray}
\label{S0}
S_i^0 (\not\! p) &=& \frac{1}{\not\! p - m_i + i\varepsilon}\ , \quad
i=t,\ t'\nonumber\\ \widehat{\Sigma}_{ij}(\not\! p) &=& \not\! p P_L\,
\Sigma^L_{ij}(s)\, +\, \not\! p P_R\, \Sigma^R_{ij}(s)\, +\, P_L\,
\Sigma^D_{ij}(s)\, +\, P_R\, \tilde{\Sigma}^D_{ij}(s)\ ,
\end{eqnarray}
with $s=p^2$. In Eq.\ (\ref{S0}), $P_{L(R)}=[1-(+)\gamma_5]/2$ is the
chirality projection operator, while the symbol hat on the individual
components of $\widehat{\Sigma}_{ij}(\not\! p)$ has been dropped for
simplicity.  Nevertheless, it must be kept in mind that the
self-energies are OS renormalized within the PT framework
\cite{JP&AP}.

Decomposing the different self-energy components into dispersive and
absorptive parts,
\begin{eqnarray}
\label{Sigma}
\Sigma^L_{ij}(s) & = & \Sigma^{L,disp}_{ij}(s)\, +\, i
                                     \Sigma^{L,abs}_{ij}(s)\, ,\qquad
\Sigma^R_{ij}(s)\ =\ \Sigma^{R,disp}_{ij}(s)\, +\, i
                                     \Sigma^{R,abs}_{ij}(s)\, ,
                                     \nonumber\\ 
\Sigma^D_{ij}(s) & = & \Sigma^{D,disp}_{ij}(s)\, +\, i
                                     \Sigma^{D,abs}_{ij}(s)\, ,\qquad
\tilde{\Sigma}^D_{ij}(s)\ =\ \Sigma^{D*,disp}_{ji}(s)\, +\, i
                                     \Sigma^{D*,abs}_{ji}(s)\, ,
\end{eqnarray}
one derives from the Hermiticity property of the Lagrangian
\begin{equation}
\label{SLdisp}
\Sigma^{L,disp}_{ij}(s)\, =\, \Sigma^{L*,disp}_{ji}(s)\, ,\quad
\Sigma^{L,abs}_{ij}(s)\, =\, \Sigma^{L*,abs}_{ji}(s)\, .
\end{equation}
Similar equalities also hold for $\Sigma^{R,disp}_{ij}(s)$ and
$\Sigma^{R,abs}_{ij}(s)$. At one loop, only the dispersive parts of
the self-energies can participate in the renormalization, since the
bare as well as counter-term (CT) Lagrangian is Hermitian
\cite{BK&AP}, whereas the absorptive self-energy parts,
$i\widehat{\Sigma}^{abs}_{ij} (\not\! p)$, are anti-Hermitian.

Another important point pertains to the issue of gauge invariance of
resonant processes involving heavy fermions, {\em e.g.}, the heavy $t$
and $t'$ quarks, after resummation has been completed. Apart from the
gauge-fixing parameter independence within the PT, one has to worry
about preserving additional gauge symmetries, when a gauge boson, such
as the $W$ or $Z$ bosons, couples to the fermionic line. In the
conventional formalism, these gauge symmetries may get distorted by
high-order quantum effects, after resummation has been carried out. In
the PT, these extra gauge symmetries are reassured by the following PT
WI's:\footnote[1]{ Similar WI's have been derived in \cite{NP} in a
semi-phenomenological manner, by studying the structure of the gauge
cancellations for different Breit-Wigner propagators. The PT WI's in
Eqs.\ (\ref{PTf1})--(\ref{PTf3}) coincide with those obtained by the
background field method in \cite{BFM}.}
\begin{eqnarray}
\label{PTf1}
p^\mu \widehat{\Gamma}_\mu^{W^+ud}(p,p_u,p_d)\, +\, M_W
\widehat{\Gamma}^{G^+ud}(p,p_u,p_d) &=&-\, \frac{ig}{\sqrt{2}}\,
V_{ud}\, \Big[\widehat{\Sigma}_{uu}(\not\! p_u) P_L\, -\, P_R
\widehat{\Sigma}_{dd}(\not\! p_d)\Big]\, ,\qquad\\
\label{PTf2}
p^\mu \widehat{\Gamma}_\mu^{W^-du}(p,p_d,p_u)\, -\, M_W
\widehat{\Gamma}^{G^-du}(p,p_d,p_u) &=&-\, \frac{ig}{\sqrt{2}}\,
V^*_{ud}\, \Big[\widehat{\Sigma}_{dd}(\not\! p_d) P_L\, -\, P_R
\widehat{\Sigma}_{uu}(\not\! p_u )\Big]\, ,\qquad\\
\label{PTf3}
p^\mu \widehat{\Gamma}_\mu^{Z\bar{f}f}(p,p_{\bar{f}},p_f)\, -\, iM_Z
\widehat{\Gamma}^{G^0\bar{f}f}(p,p_{\bar{f}},p_f) &=& \nonumber\\
&&\hspace{-180pt} \frac{ig}{2\cos\theta_w}\,
\Big[\widehat{\Sigma}_{ff}(\not\! p_{\bar{f}})
(T^f_zP_L-2Q_f\sin^2\theta_w )\, -\, (T^f_zP_R-2Q_f\sin^2\theta_w)
\widehat{\Sigma}_{ff}(\not\! p_f)\Big]\, ,\qquad
\end{eqnarray}
where $V_{ud}$ is the CKM matrix, $\theta_w$ is the weak mixing angle,
$Q_f$ is the fractional charge of quarks ({\em i.e.}, $Q_u=2/3$,
$Q_d=-1/3$) and $T^f_z$ is the $z$-component of the weak isospin of
the fermion $f$, defined after Eq.\ (\ref{chis}). In Eqs.\
(\ref{PTf1})--(\ref{PTf3}), the momentum $p_\mu$ of the gauge bosons
flows into the vertex, while the four-momenta of the fermions point to
the same direction with the fermion-number arrow. The PT WI's
(\ref{PTf1}) and (\ref{PTf2}) may be of interest here, since the PT WI
in Eq.\ (\ref{PTf3}) is only crucial when heavy fermions are produced
via $s$-channel $Z$-boson interactions.  It is obvious that possible
absorptive parts in the vertices, {\em e.g.}, $Wtb$ and $Wt'b$, will
now communicate with the respective absorptive parts of the
self-energies by means of the PT WI's (\ref{PTf1}) and (\ref{PTf2}).
Nevertheless, we can simplify things if we are only interested in
heavy fermion mass effects and so approximate the $W^\pm$ boson with
the would-be Goldstone boson $G^\pm$, as is dictated by the
equivalence theorem \cite{EqTh}. In this high-energy approximation,
it is then easy to convince oneself that there are no virtual $G^\pm$ 
corrections to the one-loop couplings $\widehat{\Gamma}^{G^+td}$ 
and $\widehat{\Gamma}^{G^+t'd}$ for $m_d=0$, because of charge
conservation on the fermionic vertices in the loop. Clearly,
this shows that the absorptive parts of the vertices are sub-dominant
in powers of $gm_t/M_W$ or $gm_{t'}/M_W$. Therefore, we think that
considering the afore-mentioned limit of the equivalence theorem will
not change our theoretical predictions significantly.

Even though the fermionic case may be more involved than that of the
scalar mixing due to the spinorial structure of fermions, the main
conceptual issues regarding particle mixing remain the same. Inverting
the matrix, $\widehat{S}^{-1}_{ij} (\not\!\! p)$ in Eq.\ (\ref{InvS}),
we arrive at the resummed fermionic propagators
\begin{eqnarray}
\label{Stt}
\widehat{S}_{tt}(\not\! p) &=& \Big[\not\! p\, -\, m_t\, +\,
\widehat{\Sigma}_{tt}(\not\! p)\, -\, \widehat{\Sigma}_{tt'}(\not\! p)
\frac{1}{\not\! p - m_{t'} + \widehat{\Sigma}_{t't'}(\not\! p)}
\widehat{\Sigma}_{t't}(\not\! p) \Big]^{-1}, \\
\label{St't'}
\widehat{S}_{t't'}(\not\! p) &=& \Big[\not\! p\, -\, m_{t'}\, +\,
\widehat{\Sigma}_{t't'}(\not\! p)\, -\, \widehat{\Sigma}_{t't}(\not\!
p) \frac{1}{\not\! p - m_t + \widehat{\Sigma}_{tt}(\not\! p)}
\widehat{\Sigma}_{tt'}(\not\! p) \Big]^{-1}, \\
\label{Stt'}
\widehat{S}_{tt'}(\not\! p) &=& -\, \widehat{S}_{tt}(\not\! p)\,
\widehat{\Sigma}_{tt'}(\not\! p)\, \Big[ \not\! p\, -\, m_{t'}\, +\,
\widehat{\Sigma}_{t't'}(\not\! p) \Big]^{-1} \nonumber\\ &=& -\, \Big[
\not\! p\, -\, m_t\, +\, \widehat{\Sigma}_{tt}(\not\! p) \Big]^{-1}
\widehat{\Sigma}_{tt'}(\not\! p)\, \widehat{S}_{t't'}(\not\! p)\ ,\\
\label{St't}
\widehat{S}_{t't}(\not\! p) &=& -\, \widehat{S}_{t't'}(\not\! p)\,
\widehat{\Sigma}_{t't}(\not\! p)\, \Big[ \not\! p\, -\, m_t\, +\,
\widehat{\Sigma}_{tt}(\not\! p) \Big]^{-1}\, \nonumber\\ &=& -\, \Big[
\not\! p\, -\, m_{t'}\, +\, \widehat{\Sigma}_{t't'}(\not\! p)
\Big]^{-1} \widehat{\Sigma}_{t't}(\not\! p)\, \widehat{S}_{tt}(\not\!
p)\ .
\end{eqnarray}

As has been mentioned above, it may be difficult to find a rotational
invariant weak basis that quantifies the magnitude of CP violation
coming entirely from $tt'$ mixing, compared to the case of two neutral
scalars.  One possibility could therefore be to factor out mass-matrix
and CKM-type mixing elements from the vertices $V^d_i$ and $V^s_i$,
and re-absorb them into the definition of effective
Hamiltonian. Another way, and perhaps the most consistent one, is to
consider the complete production and decay amplitudes at the tree
level, since $\varepsilon$- and tree-level $\varepsilon'$-type effects
may not be easily separated in most natural extensions of the SM.
Following the latter, we may define the flavour-dependent CP-violating
parameter
\begin{equation}
\label{delta}
\delta_{ds}\ =\ \left|\frac{{\cal T}^{res}_{ds}}{ \overline{\cal
T}^{res}_{ds}}\right| \ =\ \left|\frac{V^0_{is}\ [ \not\! p\, -\,
{\cal H}(\not\! p )]^{-1}_{ij}\ V^{0*}_{jd}}{V^{0*}_{is}\ [ \not\! p\,
-\, \overline{{\cal H}}(\not\! p )]^{-1}_{ij}\ V^0_{jd}} \right|,
\end{equation}
where $\overline{{\cal H}}(\not\! p )={\cal H}^T(\not\! p )$, and
$V^0_{id}$ and $V^0_{is}$ are the respective couplings $\phi^+dj$ and
$\phi^+sj$ ($j=t,t'$) in the Born approximation. Clearly, instead of
$d$ and $s$ quarks, one could consider another pair of down-type
quarks as asymptotic states, {\em e.g.} $s$ and $b$ or $b$ and
$b'$. Because of fermion-number conservation, the production, mixing,
and decay phenomena in the up-quark family will unavoidably become
manifest in the down-quark sector.

It may be worth mentioning that the diagonal elements of the
flavour-matrix $\delta$ satisfy the requirement
\begin{equation}
\label{delCPT}
\delta_{d_id_i}\ =\ 1\,
\end{equation}
as a result of CPT invariance. Thus, we expect effects of CP violation
through mixing only from off-diagonal transitions. If an observable is
sensitive to the handedness or the helicity of the asymptotic quarks,
one is then able to define CP-violating mixing parameters, such as
$\delta_{d_{iL}d_{iR}}$, which may generally deviate from unity and
consistently respect CPT invariance.  Furthermore, necessary
conditions for resonant CP violation may be derived from Eq.\
(\ref{delta}). We will see a specific example in the next section.

Finally, we briefly comment on the case of transitions that involve
heavy Majorana fermions as intermediate states. In particular, a heavy
Majorana neutrino can decay into a charged lepton, $l$, as well as
into an anti-lepton $l^C$ with the emission of a charged Higgs. Such
scenarios have received much attention, since they can account for the
baryon asymmetry in the universe \cite{BAU}. Within our formulation of
CP violation, it is evident that one could still have
\begin{equation}
\delta_{ll^C}\ =\ \delta^{-1}_{l^Cl}\ \not =\ 1\, ,
\end{equation}
in agreement with CPT invariance. Detailed study of the latter may be
given elsewhere.  

\setcounter{equation}{0}
\section{\label{sec:CPtt'} Resonant CP violation via a {\boldmath
$tt'$}  mixing}

New-physics scenarios that give rise to a CP-asymmetric $tt'$ mixing
should extend the fermionic sector of the SM in a non-trivial manner.
For instance, adding one sequential weak isodoublet, $(t',b')_L$, and
two right-handed weak iso-singlets, $t'_R$ and $b'_R$, appears to be
the most straightforward way to accomplish that purpose. In Ref.\ 
\cite{AP1}, we have analyzed CP-violating effects originating from
$tt'$ transitions in four-generation extensions with one, two and
three Higgs doublets. Unlike \cite{AP1}, we shall study the
phenomenological implications of a strong $tt'$ mixing related to the
kinematic range $m_t\approx m_{t'}$ for the LHC. Since direct
experimental searches at Tevatron find that $b'$ should be quite
heavy, $m_{b'}>85$ GeV \cite{PDG}, absorptive phases related with the
opening of intermediate decay modes, such as $t\to W^+b'$ or $t'\to
W^+b'$, are therefore suppressed.  The CP asymmetries strongly depend
on these $b'$-dependent absorptive phases, thus rendering CP violation
difficult to detect in these models.  In order to demonstrate that
resonant CP violation via fermionic mixing can still take place in the
top quark sector, we shall consider a simple CP-violating new-physics
model, in which a mirror iso-doublet is added to the field content of
the SM, apart from the sequential doublet mentioned above.

The model of our interest extends the third family of quarks of the SM
in the following way:
\begin{equation}
\label{content}
U_{1L}=\left( \begin{array}{c} t_1 \\ b_1 \end{array} \right)_L\,
,\qquad U_{2L}=\left( \begin{array}{c} t_2 \\ b_2 \end{array}
\right)_L\, ,\qquad U^C_{2L}=\left( \begin{array}{c} b^C_2 \\ t^C_2
\end{array} \right)_L\, ,\qquad t_{1R}\, ,\qquad b_{1R}\, ,
\end{equation}
with hypercharge assignments $Y(U_{1L}) = Y(U_{2L}) = 1/3$,
$Y(U^C_{2L})=-1/3$, $Y(t_{1R})=4/3$, $Y(b_{1R})=-2/3$ and weak
isospins $T(U_{1L}) =T(U_{2L})=T(U^C_{2L})=1/2$,
$T(t_{1R})=T(b_{1R})=0$. Since the only difference relative to the SM
is the addition of two weak iso-doublets with opposite hypercharges,
the above scenario is anomaly free as well; one does not need to
include extra charged lepton fields. Such new-physics extensions may
even be motivated by E$_6$ unified models \cite{Slansky,TGR}.  After
SSB, the Yukawa sector of the model reads
\begin{eqnarray}
\label{LYmir}
-{\cal L}_Y^{mass}& =& \langle \Phi^0 \rangle\, [ f^t_1\, \bar{t}_{1L}
t_{1R}\, +\,f^t_2\, \bar{t}_{2L} t_{1R} ]\, +\, \langle
\overline{\Phi}^0 \rangle\, [ f^b_1\, \bar{b}_{1L} b_{1R}\, +\,f^b_2\,
\bar{b}_{2L} b_{1R} ]\nonumber\\ &&+\, M_1 (U^C_{2L})^T i\tau_2
C^{-1}U_{1L}\, +\, M_2 (U^C_{2L})^T i\tau_2 C^{-1}U_{2L}\ +\
\mbox{H.c.}\, ,
\end{eqnarray}
with $i\tau_2$ representing the Levi-Civita antisymmetric tensor
$\varepsilon_{ij}$ and $\langle \Phi^0 \rangle=\langle
\overline{\Phi}^0 \rangle$ being the VEV of the SM Higgs doublet. If
we denote with $ m^t_{1,2} = f^t_{1,2}\langle \Phi^0 \rangle$ and
$m^b_{1,2} = f^b_{1,2}\langle \overline{\Phi}^0 \rangle$, the
$t_1t_2$- and $b_1b_2$-mass matrices are respectively given by
\begin{equation}
\label{MtMb}
M^t\ =\ \left[ \begin{array}{cc} m^t_1 & M^*_1 \\ m^t_2 & M^*_2
\end{array} \right]\, ,\qquad M^b\ =\ \left[ \begin{array}{cc} m^b_1 &
- M^*_1 \\ m^b_2 & - M^*_2 \end{array} \right]\, .\qquad
\end{equation}
The physical states for the $b$-quark sector, $b,\ b'$, and the
$t$-quark system, $t,\ t'$, are obtained through the bi-unitary
transformations
\begin{equation}
\label{Mdiag}
U^{b\dagger}_L M^b U^b_R\ =\ \widehat{M}^b\, ,\qquad U^{t\dagger}_L
M^t U^t_R\ =\ \widehat{M}^t\, ,
\end{equation}
with
\begin{equation}
\label{t_b}
\left( \begin{array}{c} b_1\\ b_2\end{array} \right)_{L,R}\ =\
U^b_{L,R}\, \left( \begin{array}{c} b\\ b'\end{array} \right)_{L,R}\,
,\qquad \left( \begin{array}{c} t_1\\ t_2\end{array} \right)_{L,R}\ =\
U^t_{L,R}\, \left( \begin{array}{c} t\\ t'\end{array} \right)_{L,R}\,
,
\end{equation}
where $U^b_{L,R}$ and $U^t_{L,R}$ are two-dimensional unitary
matrices.

We are now in a position to write down the Lagrangian, ${\cal L}_W$,
for the charged current interactions in this sequential mirror fermion
model.  The Lagrangian ${\cal L}_W$ is given by
\begin{eqnarray}
\label{LW}
{\cal L}_W &=& -\, \frac{g}{\sqrt{2}}\, W^{+\mu}\, \Big(
\bar{t}_{1L}\gamma_\mu b_{1L}\ +\ \bar{t}_{2L}\gamma_\mu b_{2L}\ +\
\bar{b}^C_{2L} \gamma_\mu t^C_{2L} \Big)\ +\ \mbox{H.c.}\nonumber\\
&=& -\, \frac{g}{\sqrt{2}}\, W^{+\mu}\, \Big(\bar{t},\ \bar{t}'\Big)\
\Big( V^L_{ij} \gamma_\mu P_L\ +\ V^R_{ij} \gamma_\mu P_R \Big) \left(
\begin{array}{c} b\\ b' \end{array} \right)\ +\ \mbox{H.c.}\, ,
\end{eqnarray}
with the mixing matrices
\begin{equation}
\label{VLVR}
V^L_{ij}\ =\ (U^{t\dagger}_LU^b_L)_{ij}\, ,\qquad V^R_{ij}\ =\ -\,
(U^t_R)^*_{2i} (U^b_R)_{2j}\, .
\end{equation}
In the last equality of Eq.\ (\ref{LW}), we have used the property
$\bar{b}^C_{2L} \gamma_\mu t^C_{2L}=-\bar{t}_{2R}\gamma_\mu b_{2R}$.
From Eq.\ (\ref{VLVR}), we see that $V^L$ is a unitary matrix but
$V^R$ is not.  These mixing matrices can mediate CP violation. In
general, there are many rephasing-invariant CP-odd quantities in this
model, which can be formed as follows:
\begin{equation}
\Im m (V^L_{tb}V^{R*}_{tb})\, ,\qquad
\Im m (V^L_{tb}V^{L*}_{t'b}V^{R*}_{tb}V^R_{t'b})\, ,\qquad \Im m
(V^L_{tb'}V^{L*}_{t'b'}V^{R*}_{tb'}V^R_{t'b'})\, , \qquad \mbox{etc.}
\end{equation}
Note that this scenario is different from a usual two-generation
mixing model in the SM, in which judicious phase rotations of the
left- and right-handed chiral fermions can be used to eliminate all
trivial CP-odd phases, thus giving rise to a real mixing matrix.  For
instance, phase re-definitions of the left-handed weak states,
$t_{1L}$, $b_{1L}$, $t_{2L}$ and $b_{2L}$, will depend on those of the
right-handed states, $b^C_{2L}$ and $t^C_{2L}$. The reason is that
such field re-phasings will simultaneously affect the gauge
interactions of the $W$ boson with the heavy quarks in Eq.\ (\ref{LW})
and the positivity of the diagonal mass matrices $\widehat{M}^b$ and
$\widehat{M}^t$.

To avoid the tight phenomenological limits, we have implicitly assumed that 
the heavy quarks $t_1$, $t_2$, $b_1$ and $b_2$ couple feebly to the two 
lighter families, {\em i.e.}, to $u$, $d$, $s$, $c$ quarks. The most 
significant constraint arises from the $Zb_R\bar{b}_R$ coupling, which may 
affect the longitudinal polarization asymmetry of the produced $b$ quarks 
measured at the CERN Large $e^+e^-$ Collider (LEP1) \cite{KPT}. The
present experimental sensitivity can only give the upper bound
$|(U^b_R)_{21}|^2 < 0.10$, which is not very restrictive. In fact, mirror
fermion models may be appealing scenarios, since they can produce a
positive shift to the observable $R_b=\Gamma (Z\to b\bar{b})/\Gamma(Z\to
\mbox{hadrons})$ in agreement with experiments at the LEP1 \cite{BBCLN}.
\begin{center}
\begin{picture}(360,120)(0,0)
\SetWidth{0.8}

\ArrowLine(0,50)(28,50)\Text(0,41)[l]{$t$}
\ArrowLine(28,50)(112,50)\Text(70,41)[]{$b,b'$}
\ArrowLine(112,50)(140,50)\Text(135,41)[r]{$t'$}
\PhotonArc(70,50)(42,0,180){3}{10}\Text(70,100)[b]{$W^+$}

\Text(70,0)[]{{\bf (a)}}


\ArrowLine(220,50)(248,50)\Text(220,41)[l]{$t$}
\ArrowLine(248,50)(332,50)\Text(290,41)[]{$b,b'$}
\ArrowLine(332,50)(360,50)\Text(355,41)[r]{$t'$}
\DashArrowArcn(290,50)(42,180,0){5}\Text(290,100)[b]{$G^+$}

\Text(290,0)[]{{\bf (b)}}

\end{picture}\\[0.7cm]
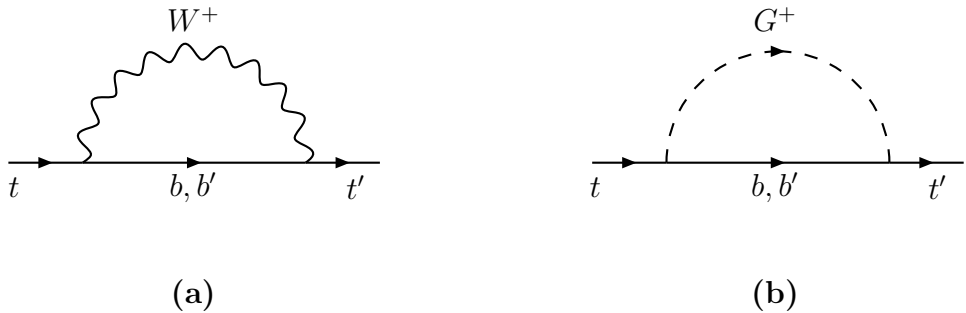
{\small {\bf Fig.\ 10:} $tt'$ transitions in models with sequential doublets.}
\end{center}

In models with additional sequential doublets, one can generally induce a
$tt'$ transition amplitude as shown in Fig.\ 10. It is therefore useful to
know the absorptive parts of all $tt$, $tt'$ and $t't'$ quark
self-energies within the PT. The best way to evaluate the PT self-energies
is to choose a gauge, in which there are no pinching $k_\mu k_\nu$ momenta
coming from the virtual $W$ propagator \cite{JMC,Yannis}. Consequently, we
shall adopt the Feynman-'t Hooft gauge, {\em i.e.}, $\xi=1$ in $R_\xi$
gauges. To avoid large effects in the electroweak oblique parameters, we
assume the mass pattern $m_{b'}\approx m_t\approx m_{t'}$, which leads to 
vanishing $b'$-absorptive contributions. Taking the above assumptions into
account, the analytic results of the quark self-energies may be cast into
the form
\begin{eqnarray}
\label{absW}
\widehat{\Sigma}^{abs}_{ij}(\not\! p) \Big|_{(a)} &=&
\frac{\alpha_w}{8}\, \Big(\, 1-\frac{M^2_W}{p^2}\, \Big)^2\, 
(V^L_{ib}V^{L*}_{jb}\, \not\! p P_L\ +\ V^R_{ib}V^{R*}_{jb}\, \not\! p P_R)
       \, ,\\
\label{absG}
\widehat{\Sigma}^{abs}_{ij}(\not\! p) \Big|_{(b)} &=&
\frac{\alpha_w}{16}\, \frac{m_i m_j}{M^2_W}\, \Big(\, 1-\frac{M^2_W}{p^2}\, 
\Big)^2\, (V^L_{ib}V^{L*}_{jb}\, \not\! p P_R\ +\ V^R_{ib}V^{R*}_{jb}\, \not\! 
p P_L) \, ,
\end{eqnarray}
where the indices $i,j$ run over $t,t'$. As can be seen from Eqs.\
(\ref{absW}) and (\ref{absG}), the self-energy graph in Fig.\ 10(a) has
different chirality structure from that shown in Fig.\ 10(b). The main
reason is that the longitudinal and transverse degrees of a gauge boson
give rise to different chirality flips. To be precise, the longitudinal
$W$ boson behaves as a scalar current which will change a left-handed
chiral fermion into a right-handed one and vice versa, whereas the
transverse component of the $W$ boson will not. In the SM with an
effective two-generation mixing, the respective analytic results may be
easily recovered by dropping all those terms that are proportional to
$V^R_{ij}$  in Eqs.\ (\ref{absW}) and (\ref{absG}).

To make our points on resonant CP violation more explicit, we shall consider 
the simple $2\to 2$ scattering process $\phi^+ s\to \phi^+ b$, which involves 
$tt'$ mixing as displayed in Fig.\ 9. Since the dominant contribution to
such a process will originate from longitudinal $W$ bosons, we will approximate
the $W^+$ boson with its unphysical would-be Goldstone boson, denoted here
as $\phi^+$. In two-Higgs doublet models or supersymmetrized versions of 
this model, $\phi^+$ could also be a physical charged Higgs boson. For 
our illustrations, we shall also assume that only the top quark couples
to the $s$ quark, {\em i.e.}, $V^L_{t's}=V^R_{t's}=0$, and the c.m.\ energy
is somehow tuned to $m_t$. Employing the relations (\ref{Stt'}) and 
(\ref{St't}) between the resummed propagators $\widehat{S}_{tt}(\not\! p)$ and 
$\widehat{S}_{t't}(\not\! p)$, $\widehat{S}_{tt'}(\not\! p)$, we may 
conveniently factorize the matrix elements of our resonant process and its 
CP-conjugate counterpart as follows:
\begin{eqnarray}
\label{Tbsfact}
{\cal T}^{res}_{sb}(\phi^+ s\to \phi^+ b) &\approx&
\frac{1}{im_t\Gamma_t}\, {\cal T}^0 (\phi^+ s\to t)\, {\cal T} (t\to \phi^+ b)
\, ,\nonumber\\ 
\overline{{\cal T}}^{res}_{sb}(\phi^+ s\to \phi^+ b) &\approx&
\frac{1}{im_t\Gamma_t}\, {\cal T}^0 (\phi^- \bar{s}\to \bar{t})\, 
{\cal T} (\bar{t}\to \phi^- \bar{b})\, ,
\end{eqnarray}
where ${\cal T}^0$ denotes the tree-level amplitude. In Eq.\ (\ref{Tbsfact}), 
$\Gamma_t$ takes account of width effects near the top-quark production. Even 
though one can always work with the exact propagator expressions, the above 
factorization of the resonant amplitudes will not significantly affect our 
quantitative discussion as far as the phenomenon of CP violation is concerned.
To make this obvious, we define the CP asymmetry
\begin{equation}
\label{aCPbs}
a_{CP}\ =\ \frac{|{\cal T}^{res}_{sb}|^2-
|\overline{{\cal T}}^{res}_{sb}|^2}{|{\cal T}^{res}_{sb}|^2+
|\overline{{\cal T}}^{res}_{sb}|^2}\ =\ \frac{\delta_{sb}^2-1}{1
+\delta_{sb}^2}\ \approx\ \frac{|{\cal T} (t\to \phi^+ b)|^2-
|{\cal T} (\bar{t}\to \phi^- \bar{b})|^2}{|{\cal T} (t\to \phi^+ b)|^2+
|{\cal T} (\bar{t}\to \phi^- \bar{b})|^2}\ .
\end{equation}
As may be seen from Eq.\ (\ref{aCPbs}), the main advantage of the above
simplifications is that the CP asymmetry $a_{CP}$ reduces to calculating
CP violation in the partial decay rate of the top quark into the charged
scalar $\phi^+$ and the $b$ quark. Up to overall coupling constants,
the matrix element for $t\to \phi^+ b$ may then be given by
\begin{eqnarray}
\label{Tphib}
{\cal T} (t\to \phi^+ b) &\sim & \frac{m_t}{M_W}\, 
\bar{u}_b\Big( V^{L*}_{tb} P_R\, +\, V^{R*}_{tb} P_L\, \Big) u_t\
-\ i\, \frac{m_{t'}}{M_W}\, \bar{u}_b \Big( V^{L*}_{t'b} P_R\,
+\, V^{R*}_{t'b} P_L\, \Big) \nonumber\\
&&\times\, 
\Big[ \not\! p - m_{t'} + i\widehat{\Sigma}^{abs}_{t't'}(\not\! p)\Big]^{-1} 
\widehat{\Sigma}^{abs}_{t't}(\not\! p)\, u_t\, .
\end{eqnarray}
Similarly, the CP-conjugate decay matrix element may be written down 
\begin{eqnarray}
\label{TCPphib}
\overline{{\cal T}} (\bar{t}\to \phi^- \bar{b}) &\sim & \frac{m_t}{M_W}\, 
\bar{v}_t\Big( V^L_{tb} P_L\, +\, V^R_{tb} P_R\, \Big) v_b\
-\ i\, \frac{m_{t'}}{M_W}\, \bar{v}_t 
\widehat{\Sigma}^{abs}_{t't}(-\not\! p)  \nonumber\\
&&\times\, 
\Big[ -\not\! p - m_{t'}+i\widehat{\Sigma}^{abs}_{t't'}(-\not\! p)\Big]^{-1} 
\, \Big( V^L_{t'b} P_L\,
+\, V^R_{t'b} P_R\, \Big) v_b\nonumber\\
&=&\frac{m_t}{M_W}\, 
\bar{u}_b\Big( V^L_{tb} P_L\, +\, V^R_{tb} P_R\, \Big) u_t\
-\ i\, \frac{m_{t'}}{M_W}\, \bar{u}_b \Big( V^L_{t'b} P_L\,
+\, V^R_{t'b} P_R\, \Big) \nonumber\\
&&\times\, 
\Big[ \not\! p - m_{t'} + i\widehat{\Sigma}^{abs,C}_{t't'}(\not\! p)\Big]^{-1} 
\widehat{\Sigma}^{abs,C}_{tt'}(\not\! p)\, u_t\, .
\end{eqnarray}
In the derivation of the last step of Eq.\ (\ref{TCPphib}), we have used
the known identities: $u(p,s)=C\bar{v}^T(p,s)$ and $C\gamma_\mu C^{-1}
=-\gamma_\mu^T$. If we now define the absorptive $j\to i$ self-energies 
($i,j=t,t'$) as
\begin{equation}
\label{defS}
\widehat{\Sigma}^{abs}_{ij} (\not\! p)\ =\ A^L_{ij} \not\! p P_R\,
+\, A^R_{ij} \not\! p P_L\, ,
\end{equation}
with $A^L_{ij}=V^L_{ib}V^{L*}_{jb} A_{ij}$, 
$A^R_{ij}=V^R_{ib}V^{R*}_{jb} A_{ij}$ and
\begin{equation}
\label{Aij}
A_{ij}\ =\ \frac{\alpha_w}{16}\, \frac{m_im_j}{M^2_W}\,
\Big(\, 1-\frac{M^2_W}{p^2}\, \Big)^2\, ,
\end{equation}
then the absorptive part of the charge-transform transition, $\bar{j}\to
\bar{i}$, is given by
\begin{equation}
\label{defSC}
\widehat{\Sigma}^{abs,C}_{ij} (\not\! p)\ =\ A^L_{ij} \not\! p P_L\,
+\, A^R_{ij} \not\! p P_R\ .
\end{equation}
Comparing Eq.\ (\ref{Tphib}) with Eq.\ (\ref{TCPphib}), it is easy to
observe that, under a CP transformation, the mixing matrices $V^L$, $V^R$ 
and the chirality projector $P_L$ are mapped into the mixing matrices 
$V^{L*}$, $V^{R*}$ and $P_R$, respectively, while all the absorptive
phases remain unchanged. By virtue of the Dirac equation of motion,
the matrix element ${\cal T} (t\to \phi^+ b)$ simplifies to the expression
\begin{equation}
\label{TLR}
{\cal T} (t\to \phi^+ b)\ =\  {\cal T}_L (t\to \phi^+ b)\ +\ 
{\cal T}_R (t\to \phi^+ b)\, ,
\end{equation}
where 
\begin{eqnarray}
\label{TL}
{\cal T}_L (t\to \phi^+ b) &\sim& \bar{u}_b P_L u_t\nonumber\\
&&\hspace{-20pt} \times\, 
\left[\, V^{R*}_{tb}\, \frac{m_t}{M_W}\ -\ 
iV^{R*}_{t'b}\, \frac{m_{t'}}{M_W}\, 
\frac{m^2_t (1+iA^R_{t't'})A^R_{t't} + m_t m_{t'}A^L_{t't}}{
m^2_t (1+iA^L_{t't'})(1+iA^R_{t't'}) - m^2_{t'} }\, \right]
\end{eqnarray}
and 
\begin{eqnarray}
\label{TR}
{\cal T}_R (t\to \phi^+ b) &\sim& \bar{u}_b P_R u_t\nonumber\\
&&\hspace{-20pt} \times\, 
\left[\, V^{L*}_{tb}\, \frac{m_t}{M_W}\ -\ 
iV^{L*}_{t'b}\, \frac{m_{t'}}{M_W}\, 
\frac{m^2_t (1+iA^L_{t't'})A^L_{t't} + m_t m_{t'}A^R_{t't}}{
m^2_t (1+iA^L_{t't'})(1+iA^R_{t't'}) - m^2_{t'} }\, \right] .
\end{eqnarray}
Neglecting small $m_b$-dependent terms, one can now verify that the
CP-violating contributions of $|{\cal T}_L|^2$ and $|{\cal T}_R|^2$ to
$a_{CP}$ defined in Eq.\ (\ref{aCPbs}) cancel in the sum.  The
underlying reason responsible for such a cancellation is the
invariance of the process under CPT transformations.  Evidently, if
there was an extra charged Higgs, such as $\phi^+$, which preferred to
couple to right-handed $b$ quarks and/or left-handed top quarks only,
then $a_{CP}$ would not vanish.

Motivated by the recent studies of probing CP violation in the production
of polarized top decays at the LHC \cite{Peskin,WB}, we shall focus our
attention on the decay $t_L\to \phi^+b$ and the CP-conjugate decay
$\bar{t}_R\to \phi^- \bar{b}$. In our prototype scattering $\phi^+ s\to
\phi^+ b$, this would amount to assuming that the $s$ quark couples to 
the left-handed top quark only. The matrix element for the decay 
$t_L\to \phi^+ b$ may be obtained from ${\cal T}_L (t\to \phi^+ b)$ in
Eq.\ (\ref{TL}). In this way, we find for the CP asymmetry $A^{pol}_{CP}$,
\begin{equation}
\label{Apol}
A^{pol}_{CP} \ =\ \frac{\Gamma (t_L\to \phi^+ b) -
\Gamma (\bar{t}_R\to \phi^- \bar{b}) }{
\Gamma (t_L\to \phi^+ b) + \Gamma (\bar{t}_R\to \phi^- \bar{b})}\ .
\end{equation}
By analogy, one could define the CP asymmetry $\overline{A}^{pol}_{CP}$
based on the decays $t_R\to \phi^+b$ and the CP transforms, $\bar{t}_L\to 
\phi^-\bar{b}$. Since our primary interest lies in the region for resonant 
CP-violating $tt'$ transitions, we may simplify calculation by making
the approximations $\Delta m^2_t = m^2_t - m^2_{t'} \ll m^2_t,\ m^2_{t'}$.
Defining the ratio $r_t = \Delta m^2_t/m^2_t$ and neglecting terms of
order $A^2_{ij}$ where possible, we derive the simple expression for the 
CP asymmetry
\begin{equation}
\label{ApolCP}
A^{pol}_{CP}\ \approx\  
-\, \frac{2\Im m (V^L_{tb}V^{L*}_{t'b}V^{R*}_{tb}V^R_{t'b})}{
|V^R_{tb}|^2}\ \frac{r_t A_{tt}}{r^2_t\, +\, 
( |V^R_{t'b}|^2 + |V^L_{t'b}|^2 )^2 A^2_{tt}}\ .
\end{equation}
The analytic form for $\overline{A}^{pol}_{CP}$ may also be obtained from
Eq.\ (\ref{ApolCP}), if one replaces  $|V^R_{tb}|^2$ with $|V^L_{tb}|^2$
and changes the overall sign. In Fig.\ 11, we have plotted the dependence
of $A^{pol}_{CP}$ as a function of $r_t$ for different values of the mixing 
matrix combination $|V^R_{t'b}|^2+|V^L_{t'b}|^2$. The CP asymmetry
$\overline{A}^{pol}_{CP}$ is given in units of $\Im m 
(V^L_{tb}V^{L*}_{t'b}V^{R*}_{tb}V^R_{t'b})/|V^R_{tb}|^2$, which is generally 
smaller than 0.5. As can be seen from Fig.\ 11, we recover the known feature 
of resonant CP violation through particle mixing, namely, CP violation may 
become maximal for small values of the parameter $r_t$, {\em i.e.},
$10^{-4}< r_t< 10^{-1}$.  

This mixing mechanism, through which CP violation is resonantly
amplified, may take place in scatterings where top quarks are produced
either singly or in pairs via gluon fusion processes. In fact, one
expects to be able to analyze about $10^6-10^7$ top decays at the LHC
for an integrated luminosity of 100 fb$^{-1}$. Obviously, various
techniques in analyzing top quark polarization have been suggested in
the literature \cite{Nacht,Peskin,WB}, which may help to study
resonant CP-violating effects at high-energy colliders. As has been
shown in this section, CP violation may be of order one due to $tt'$
resonant transitions, especially when the mass difference $m_t-m_{t'}$
lies in the vicinity of the widths of the $t$ and $t'$ quarks. This
gives a unique chance to probe such effects in future experiments and
so unravel the underlying CP nature of the top-quark sector.

\setcounter{equation}{0}
\section{\label{sec:Concl} Conclusions}

The CP-violating dynamics known from the $K^0\bar{K}^0$ system has been
extended to processes that can take place at high-energy colliders. At low
energies, one may carry out measurements based on the time evolution of
the unstable kaons, whereas, at high energies, one is compelled to consider
reactions that can only be described by scattering amplitudes. 
This constitutes a non-trivial generalization of the $K^0\bar{K}^0$ dynamics.
In transition amplitudes, the r\^ole of time assumes its Fourier-conjugate
variable, the energy.  At high energies, it is therefore crucial
to study the dependence of CP violation as a function of invariant mass
energies and/or momenta of the asymptotic final states, such as charged 
leptons and jets.

As has been analyzed in Sections \ref{sec:Bos} and \ref{sec:Fer},
CP-violating phenomena can be significantly enhanced through the
mixing of two resonant particles that behave differently under CP and
whose mass difference is comparable to their widths. In particular,
the underlying mechanism for large CP violation induced by resonant
bosonic as well as fermionic transitions has been clarified and
studied carefully on a more rigorous field-theoretic basis.  In this
context, we have considered a resummation approach, which is
implemented by the PT \cite{JP&AP} and hence preserves the gauge
symmetries of the theory.

Models that may give rise to non-negligible bosonic $HA$ and/or
fermionic $tt'$ mixings are discussed in Sections \ref{sec:HA} and
\ref{sec:CPtt'}.  In Sections \ref{sec:CPHA} and \ref{sec:CPtt'}, we
have further analyzed the phenomenological implications of our
mechanism for large CP-violating phenomena in the production, mixing
and decay of a top quark or a Higgs particle at planned high-energy
machines, such as the LHC, NLC and/or muon collider. Since high order
$\varepsilon'$-type effects are generally suppressed near the resonant
region, possible large CP-violating phenomena can naturally be
accounted for by the mixing mechanism presented in this paper.

Our analysis given in Section \ref{sec:RCPV} has shown that the
CP-violating phenomenon becomes maximal, {\em i.e.}, it could be of
order one, when the two non-free particles are degenerate but possess
an {\em anomalous} non-vanishing mixing.  To the best of our
knowledge, this is a novel aspect, which has not been addressed
properly in the literature before.  Moreover, we have paid special
attention to possible constraints imposed by CPT invariance on the
actual magnitude of CP violation in Section \ref{sec:CPT}. On the
other hand, approaches based on diagonalizing the effective
Hamiltonian by means of a similarity transformation are inadequate to
deal with an {\em anomalously} degenerate mixing system. For instance,
in a $K^0\bar{K}^0$-like basis, such an anomalously degenerate mixing
system is manifested by a non-diagonalizable effective Hamiltonian of
the Jordan form. As has been demonstrated in Section \ref{sec:WW}, the
transformation matrix $X$ becomes singular in such a case.  However, our
formalism does not display this kind of singularity, since it makes
use of the well-defined properties of the transition amplitudes, for
which such similarity transformations are not needed.  Therefore, it
may be fair to say that our field-theoretic approach unifies features
of the effective Hamiltonian \cite{WW} and/or the propagator formalism
by Sachs \cite{Sachs} with those pertaining to the density matrix
\cite{DMF}.  The significance of this resonantly amplified
CP-violating mechanism for other anomalously degenerate physical
systems appears to be an open challenge for further investigations.

\bigskip
\noindent {\bf Acknowledgments.} I wish to thank Emmanuel Paschos for
enlightening conversations and for his continuous encouragement to
complete this work.  I am also very much benefited from discussions
with Pasha Kabir, Joannis Papavassiliou, Leo Stodolsky and Arkady
Veinshtein.

\newpage

\def\theequation{\Alph{section}.\arabic{equation}}
\begin{appendix}
\setcounter{equation}{0}
\section{Mixing renormalization in scalar theories}

If the $HA$ mixing occurs at the tree-level, one may then have to worry about
the appearance of UV infinities that arise generally at higher loops
\cite{Ren,Sirlin1}. Facing this problem is unavoidable, since one-loop 
absorptive corrections must be considered in the calculation, otherwise our CP
asymmetries (see, {\em e.g.}, Eq.\ (\ref{aCP1}))  will vanish identically.
Therefore, renormalization of the UV divergences in the presence of a $HA$
mixing requires particular care.  Here, we shall extend the mixing
renormalization programme, presented in \cite{BK&AP} for fermions, to the 
mixing of scalar particles.

Let us consider the kinetic part of the bare Lagrangian governing the mixing of
$N$ real scalars $S'^0_i$ ($i=1,2,\dots ,N$):
\begin{equation}
\label{bare}
{\cal L}_{kin}^0\ =\ \frac{1}{2} (\partial_\mu S'^{0T})
(\partial^\mu S'^0)\, -\, \frac{1}{2} S'^{0T} (M'^0)^2 S'^0\, . 
\end{equation}
Here and in the following, we use the convention to denote quantities
that are expressed in the flavour basis by a prime, {\em e.g.}, $S'_i$,
$M'_{ij}$, while we attach the superscript `0' for all unrenormalized 
quantities. In general, the mass matrix $M'^0$ in Eq.\ (\ref{bare}) is a
$N\times N$ dimensional real, positive semi-definite and symmetric matrix.
This matrix can then be diagonalized by performing an orthogonal rotation,
$O^0$, of the weak fields $S'^0_i$, {\em i.e.},
\begin{equation}
\label{rot0}
(M^0)^2\ =\ O^0 (M'^0)^2 O^{0T}\, ,\qquad S^0\ =\ O^0 S'^0\, ,
\end{equation} 
where the absence of a prime on the fields and the kinematic parameters
indicates that these quantities are written in the mass basis. As a result, 
the bare mass matrix $M^0$ is a non-negative diagonal matrix of $N\times N$ 
dimensions. Following \cite{BK&AP}, we introduce counter-terms (CT's) and 
express the bare quantities in terms of the renormalized ones:
\begin{eqnarray}
\label{CT1}
O^0 &=& O\, +\, \delta O\, ,\\
\label{CT2}
(M'^0)^2 &=& M'^2\, +\, \delta M'^2\, ,\\
\label{CT3}
(M^0)^2 &=& M^2\, +\, \delta M^2\, ,\\
\label{CT4}
S'^0 &=& Z'^{1/2} S'\ =\ (1\, +\, \frac{1}{2} \delta Z')\, S'\, ,\\
\label{CT5}
S^0 &=& Z^{1/2} S\ =\ (1\, +\, \frac{1}{2} \delta Z)\, S\, ,
\end{eqnarray}
where $\delta M'^2$ ($\delta M^2$) and $Z'^{1/2}$ ($Z^{1/2}$) are the 
mass and wave-function renormalization constants in the flavour (mass)
basis. Note that a novel CT for the orthogonal matrix $O$, $\delta O$, is
induced by this procedure. If we impose that unitarity, precisely speaking
orthogonality, of the theory should hold order by order in perturbation
theory for the bare as well as renormalized orthogonal matrix $O$, {\em
i.e.}, $OO^T = O^0O^{0T} = 1$, it is then easy to find that the matrix
$O\delta O^T$ is anti-orthogonal, {\em viz.}
\begin{equation}
\label{antiO}
O\delta O^T\ =\ -\delta O O^T\, .
\end{equation} 
Taking Eqs.\ (\ref{CT1})--(\ref{CT5}) into account, the one-loop CT
Lagrangian reads:
\begin{equation}
\label{CTLkin}
\delta {\cal L}_{kin}\ =\ \frac{1}{4}\Big[(\partial_\mu S^T)(\delta Z +
\delta Z^T)(\partial^\mu S)\, -\, S^T(2\delta M^2 + M^2\delta Z  
+ \delta Z^T M^2) S\Big]\, .
\end{equation}
Since both the bare and CT Lagrangians are Hermitian, it is evident that
only the dispersive parts of the two-point correlation functions
should enter the renormalization. 

We can now proceed renormalizing the one-loop transitions, $S_j\to S_i$. 
In general, these transitions are described by the unrenormalized self-energy 
functions $\Pi_{ij} (p^2)$. Considering the CT Lagrangian in Eq.\
(\ref{CTLkin}), the renormalized self-energies $\widehat{\Pi}_{ij}(p^2)$
may be written down
\begin{equation}
\label{Piren}
\widehat{\Pi}_{ij}(p^2)\ =\ \Pi_{ij}(p^2) + \frac{p^2}{2}(\delta Z_{ij}
+\delta Z_{ji})-\delta_{ij}\delta M^2_{ij}-\frac{1}{2}(M^2_i\delta Z_{ij}
+\delta Z_{ji}M^2_j)\, .
\end{equation}
Note that $\Pi_{ij}(p^2)=\Pi_{ji}(p^2)$, which also implies that
$\widehat{\Pi}_{ij}(p^2)=\widehat{\Pi}_{ji}(p^2)$. In addition, the
renormalized self-energies satisfy the following OS renormalization
conditions:
\begin{eqnarray}
\label{OS1}
\Re e\widehat{\Pi}_{ij}(M^2_j)\ =\ \Re e\widehat{\Pi}_{ji}(M^2_i)& =& 0\, ,\\
\label{OS2}
\lim\limits_{p^2\to M^2_i}\ \frac{1}{p^2-M^2_i}\, \Re e\widehat{\Pi}_{ii}(p^2)
&=& 0\, . 
\end{eqnarray}
We could also choose another renormalization scheme, in which the
self-energies are renormalized by requiring that the complex pole
positions of the matrix elements are not shifted.  Even though such a
scheme is shown to be gauge independent \cite{Sirlin2} in the weak
mixing limit, it is, however, more involved than OS renormalization,
if the mixing becomes strong. In Appendix B, we will show that the
difference between OS and pole-mass renormalization is of higher order
in the absence of particle mixing; they differ at two loops. In the
presence of a large mixing, OS renormalization constitutes a more
natural scheme, since the respective eigenvectors of the
OS-renormalized masses form an orthogonal (in general unitary) Hilbert
space, whereas the eigenvectors of the pole masses do not.
Furthermore, we calculate all the one-loop OS renormalized
self-energies within the PT framework, so as to avoid problems arising
from possible violations of gauge symmetries when resummation is
considered.

From the OS conditions in Eqs.\ (\ref{OS1}) and (\ref{OS2}), we can now
calculate the mass and wave-function CT's in terms of bare
self-energies:
\begin{eqnarray}
\label{dM}
\delta M^2_i &=& \Re e\Pi_{ii} (M^2_i)\, , \\
\label{dZii}
\delta Z_{ii} &=& -\, \Re e\Pi'_{ii} (M^2_i)\, ,\\
\label{dZij}
\delta Z_{ij} &=& \frac{2\Re e\Pi_{ij}(M^2_j)}{M^2_i-M^2_j}\, .
\end{eqnarray}
In Eq.\ (\ref{dZii}), the prime on the self-energy denotes
differentiation with respect to $p^2$ at the position $p^2=M^2_i$.
The remaining CT, $\delta O$, is indispensable when one considers
the renormalization of vertex interactions. For example, let us
assume that the scalars $S'^0_i$ couple to the fermion field $f^0$,
{\em e.g.},
\begin{equation}
\label{Lint}
{\cal L}^0_{int}\ =\ g^0 S'^0_i\, \bar{f}^0\, f^0\, ,
\end{equation}
where $g^0$ is the bare coupling constant.  As we will see in a moment,
our derivation of mixing renormalization will not depend upon other model
details, {\em e.g.}, if additional fermions with scalar and/or pseudo-scalar
couplings are present in the interaction Lagrangian (\ref{Lint}).
Expressing ${\cal L}_{kin}$ in terms of mass-eigenfields and renormalized
quantities, we obtain the CT Lagrangian for ${\cal L}_{int}$,
\begin{equation}
\label{dLint}
\delta {\cal L}_{int}\ =\ g \delta O_{ji} S_j\bar{f}f\, +\,
g O_{ki} \Big(\frac{1}{2}\delta Z_{kj} + \delta_{kj}\frac{\delta g}{g}
+\delta_{kj}\delta Z_f\Big) S_j \bar{f}f\, .
\end{equation}
Since the mixing renormalization should not depend on the coupling
constant CT, $\delta g$, and the fermion wave-function 
renormalization, $\delta Z_f$, one is then left with the fact that 
$\delta O$ must only depend on $O_{ij}$ and $\delta Z_{ij}$.
Employing Eq.\ (\ref{antiO}), we can rewrite $\delta {\cal L}_{int}$
in matrix notation as follows:
\begin{equation}
\label{Lint1}
\delta {\cal L}_{int}\ =\ g\, O^T \Big( -\delta O O^T\, +\, 
\frac{1}{2}\delta Z \Big) S \bar{f} f\, +\ \dots\, ,
\end{equation}
where the ellipses denote the other terms omitted in Eq.\ (\ref{Lint1}).
One can now take advantage of the fact that the matrix $\delta O O^T$ is 
an anti-orthogonal matrix and so re-express the latter in terms of 
the wave-function CT's as
\begin{equation}
\label{mixCT}
\delta O_{ij}\ =\ \frac{1}{4}\, (\delta Z_{il}\, -\,
\delta Z_{li} ) O_{lj}\, .
\end{equation}
In this way, the anti-symmetric parts of $\delta Z_{ij}$ are completely
absorbed by the CT's $(\delta O O^T)_{ij}$ in Eq.\ (\ref{Lint1}), whereas
the remaining symmetric CT's, $(\delta Z_{ij} + \delta Z_{ji})/4$, are
necessary for the renormalization of the one-loop irreducible $S_i\bar{f}f$ 
couplings.  As has already been pointed out in \cite{BK&AP} for the case of 
fermionic mixing which is here true as well, Eq.\ (\ref{mixCT}) is unique
up to finite anti-orthogonal terms, $\kappa_{ij}$, which quantify possible
deviations of different mixing renormalization schemes.  Nevertheless, the
scheme used here is characterized by its simplicity and may hence be
considered as quite natural.  We will not pursue this issue any further.
Instead, we remark that the above one-loop renormalization analysis for
real scalars can equally well apply to complex or charged scalars, by
making the replacements $Z^T\to Z^\dagger$ and $O^T\to O^\dagger$, where
appropriate.

\setcounter{equation}{0}
\section{Issues of mass renormalization}

In our calculations, we have used OS renormalized masses for the resonant
particles. Another physical renormalization scheme is obtained if one
requires that the resummed propagators do not shift the complex pole
positions of the S matrix.  This renormalization also yields a gauge
independent CT for the mass of the unstable particle at higher orders
\cite{Sirlin2}.  As has been shown in \cite{JP&AP} under plausible
assumptions, the very same property shares our PT resummation approach.
Apart from being gauge independent, the higher order PT self-energies do
not shift the physical complex pole \cite{JP&AP,PS,PS'}. Nevertheless, OS
renormalization is still a good renormalization framework, since it is
simpler than pole-mass renormalization. Here, we shall briefly review the
conditions of pole-mass renormalization \cite{Sirlin2} and show that the
difference between these two schemes is beyond one-loop order in the
absence of mixing. If mixing among particles is present, pole-mass
renormalization becomes more complicated. As we will see, such a 
complication may be avoided if one expresses the pole masses in terms of OS
renormalized masses and self-energies.

For our illustrations, we shall first consider just one un-mixed scalar
$S$. The inverse resummed propagator of $S$ is then given by
\begin{equation} 
\label{BD0}
\Delta^{-1} (s)\ =\ s\, -\, (M^0)^2\, +\, \Pi (s)\, ,
\end{equation}
with $s=p^2$. In the pole-mass renormalization, one makes use of the fact
that the $s$-channel exchange transition elements will exhibit a complex pole 
at the position $\bar{s} = \overline{M}^2 - i\overline{M}\overline{\Gamma}$.
In such a scheme, one imposes the physical condition
\begin{equation} 
\label{Bpole}
\bar{s}\, -\, (M^0)^2\, +\, \Pi (\bar{s})\ =\ 0\, .
\end{equation}
In this scheme, $\overline{M}$ and $\overline{\Gamma}$ are the physical
pole mass and width of the particle $S$. The pole-mass CT may be found by
writing $(M^0)^2 = \overline{M}^2 + \delta\overline{M}^2$ and substituting 
this expression into Eq.\ (\ref{Bpole}). Through order $g^4$, this leads to
\begin{eqnarray}
\label{BdM2}
\delta\overline{M}^2 &=& \Re e\Pi (\bar{s})\ =\ \Re e\Pi(\overline{M}^2)
\, +\, \overline{M}\overline{\Gamma}\Im m\Pi'(\overline{M}^2)\, ,\\
\label{BMG}
\overline{M}\overline{\Gamma} &=& \Im m \Pi (\bar{s})\ =\
\Im m\Pi(\overline{M}^2)\, -\, \overline{M}\overline{\Gamma}
\Re e\Pi'(\overline{M}^2)\, .
\end{eqnarray}
The coupled Eqs. (\ref{BdM2}) and (\ref{BMG}) may be further disentangled
perturbatively as
\begin{eqnarray}
\label{dM2}
\delta\overline{M}^2 &=& \Re e\Pi(\overline{M}^2)
\, +\, \Im m\Pi'(\overline{M}^2)\, \Im m\Pi(\overline{M}^2)\, ,\\
\label{MG}
\overline{M}\overline{\Gamma} &=& \Big[ 1 - \Re e\Pi'(\overline{M}^2)\Big]
\Im m\Pi(\overline{M}^2)\, .
\end{eqnarray}
It has been shown in \cite{Sirlin2} that the $Z$-pole mass CT
$\delta\overline{M}^2$ defined in Eq.\ (\ref{dM2}) is gauge
independent through order $g^4$. The latter should also hold true for
scalar particles present in gauge field theories. Moreover, making use
of a Taylor series expansion in Eq.\ (\ref{BdM2}), it is
straightforward to find the relation between OS renormalized mass,
$M$, and the pole mass $\overline{M}$. This relation is given by
\begin{equation}
\label{OSpole}
M^2\ = \Big[ 1\, -\, \frac{\overline{\Gamma}}{\overline{M}}\,
\Im m\widehat{\Pi}'(\overline{M}^2)\, \Big]\, \overline{M}^2\, ,
\end{equation}
where the self-energy $\widehat{\Pi}(s)$ is OS renormalized, as has been
discussed in Appendix A. Eq.\ (\ref{OSpole}) explicitly demonstrates that
the difference between the two masses enters through order $g^4$.  Similarly,
one can derive from Eq.\ (\ref{BMG}) that the OS and pole renormalized
width differ from one another at the three-loop order.

The situation is different if mixing among scalars is present. Let us
consider the mixing of two scalars, {\em e.g.}, $A$ and $H$, which give
rise to two complex poles in S-matrix elements at positions $\bar{s}_A =
\overline{M}^2_A-i\overline{M}_A\overline{\Gamma}_A$ and $\bar{s}_H = 
\overline{M}^2_H-i\overline{M}_H\overline{\Gamma}_H$. A straightforward
way to calculate the pole-mass CT's is first to write down the bare masses 
of $A$ and $H$ as
\begin{equation}
\label{CTMHA}
(M^0_A)^2\ =\ \overline{M}^2_A\, +\, \delta \overline{M}^2_A\, ,\qquad
(M^0_H)^2\ =\ \overline{M}^2_H\, +\, \delta \overline{M}^2_H\, ,
\end{equation}
and then impose the no-shift condition of the two complex poles, which 
is translated into the vanishing of the determinant of the inverse 
(unrenormalized) propagator, $\Delta^{-1}(s)$ in Eq.\ (\ref{InvDHA}), 
{\em i.e.}
\begin{equation}
\label{Rendet}
\mbox{det}\, \left[ \begin{array}{cc}
s\, -\, \overline{M}^2_A\, -\, \delta \overline{M}^2_A\, +\, \Pi_{AA}(s) & 
                         \Pi_{HA}(s) \\
\Pi_{HA}(s) &
s\, -\, \overline{M}^2_H\, -\, \delta \overline{M}^2_H\, +\, \Pi_{HH}(s)
\end{array} \right]\ =\ 0\, .
\end{equation}
Eq.\ (\ref{Rendet}) is quadratic in the variable $s$, leading to two
complex eigenvalue equations:
\begin{equation}
\label{sAH}
\bar{s}_A\, -\, \lambda_A(\bar{s}_A)\ =\ 0\, ,\qquad
\bar{s}_H\, -\, \lambda_H(\bar{s}_H)\ =\ 0\, ,
\end{equation}
In the kinematic range $s\approx \overline{M}^2_H \approx
\overline{M}^2_A$, the functions $\lambda_A (s)$ and $\lambda_H (s)$
may be approximated by the two complex mass eigenvalues of the effective
Hamiltonian for the $HA$ system ({\em cf.}\ Eq.\ (\ref{polmas})).
Retaining the full $s$-dependence of the functions $\lambda_A$ and
$\lambda_H$, we find that their explicit form is given by
\begin{eqnarray}
\label{lambdas}
\lambda_{A(H)} (s) &=& \frac{1}{2}\bigg\{\, (M^0_A)^2 + (M^0_H)^2
-\Pi_{AA} (s) -\Pi_{HH}(s) \nonumber\\
&-(+)& \Big[\Big( (M^0_A)^2 - (M^0_H)^2
-\Pi_{AA} (s) +\Pi_{HH}(s)\Big)^2\, +\, 4\Pi^2_{HA}(s)\Big]^{1/2}\, \bigg\}\, 
,\quad
\end{eqnarray}
where the bare masses of $A$ and $H$ are given in Eq.\ (\ref{CTMHA}).
Unlike Eq.\ (\ref{polmas}), Eq.\ (\ref{sAH}) is a coupled system of
two complex (or four real) equations with four unknown parameters: the
two pole-mass CT's, $\delta \overline{M}^2_A$ and $\delta
\overline{M}^2_H$, and the two pole widths,
$\overline{M}_A\overline{\Gamma}_A$ and
$\overline{M}_A\overline{\Gamma}_H$. Instead of solving this $4\times
4$ system, the best way is to renormalize the $HH$ and $AA$
self-energies in the OS scheme, by decomposing the $H$ and $A$ bare
masses into the form given in Eq.\ (\ref{CT3}). Then, Eq.\ (\ref{sAH})
together with Eq.\ (\ref{lambdas}) just display the relations between
pole masses and OS renormalized masses in the presence of a
non-vanishing particle mixing.  This is precisely the avenue that has been
followed throughout our analysis for the bosonic as well as fermionic
case.

\end{appendix}

\newpage

\centerline{\Large{\bf Captions of remaining figures}}
\vspace{-0.2cm}
\newcounter{fig}
\begin{list}{\rm {\bf Fig. \arabic{fig}: }}{\usecounter{fig}
\labelwidth1.6cm \leftmargin2.5cm \labelsep0.4cm \itemsep0ex plus0.2ex }

\item[{\bf Fig.\ 5:}] {\bf (a)} Numerical estimates of production 
cross-sections and CP violation for $\mu^-_{L,R}\, \mu^+_{L,R}\to h^*(H^*),\ 
A^* \to f\bar{f}$ as a function of c.m.\ energy $\sqrt{s}$.  {\bf (b)} CP 
asymmetry versus $x_A=\Pi^{SA}/\Im m(\widehat{\Pi}^{SS} -
\widehat{\Pi}^{AA})$, with $S=h,\ H$.

\item[{\bf Fig.\ 6:}] Numerical estimates of CP asymmetries for 
$e^+e^-\to h^* (H^*), A^*\to f_L\bar{f}_L X$ at the NLC
as a function of {\bf (a)} $f\bar{f}$ invariant mass and {\bf (b)}
$x_A=\Pi^{SA}/\Im m(\widehat{\Pi}^{SS} - \widehat{\Pi}^{AA})$,
with $S=h,\ H$. 

\item[{\bf Fig.\ 11:}] Numerical estimates of CP asymmetries in the 
production of polarized top quarks at the LHC as a function of
the ratio $r_t=(m^2_t-m^2_{t'})/m^2_t$ for different values of
the mixing matrix combination $|V^R_{t'b}|^2+|V^L_{t'b}|^2= 1.0,\
0.5,\ 0.1$. 

\end{list}
\newpage

\begin{figure}[ht]
   \leavevmode
 \begin{center}
   \epsfxsize=17.3cm
   \epsffile[0 0 539 652]{rcpvfig5.eps}
 \end{center}
\end{figure}

\newpage
\begin{figure}[ht]
   \leavevmode
 \begin{center}
   \epsfxsize=17.3cm
   \epsffile[0 0 539 652]{rcpvfig6.eps}
 \end{center}
\end{figure}

\newpage
\begin{figure}[ht]
   \leavevmode
 \begin{center}
   \epsfxsize=15.0cm
   \epsffile[0 0 425 425]{rcpvfig11.eps}
 \end{center}
\end{figure}

\end{document}